\documentclass[authoryear,preprint,12pt]{elsarticle}

\usepackage{hyperref}
\usepackage{algorithmicx}
\usepackage{algpseudocode}
\usepackage{algorithm}
\usepackage{subcaption}
\usepackage{xcolor}

\usepackage{amssymb}
\usepackage{amsmath}
\journal{Expert Systems with Applications}

\begin{document}

\begin{frontmatter}

\title{From Occasional to Steady: Habit Formation Insights From a Comprehensive Fitness Study}

%\author{Author details removed for double-blind submission}

\author[1,4]{Ege Demirci}

\author[1,2,5]{Efe Tüzün}

\author[1,2,6]{Ahmet Furkan Ün}

\author[1,2,7]{Taner Giray Sönmez}

\author[1,3]{Onur Varol\corref{cor1}}
\ead{onurvarol@sabanciuniv.edu}

\cortext[cor1]{Corresponding author}

\affiliation[1]{organization={Faculty of Engineering and Natural Sciences}, addressline={Sabanci University}, city={Istanbul}, postcode={34956}, country={Türkiye}}
\affiliation[2]{organization={Mars Athletic Club}, city={Istanbul}, postcode={34340}, country={Türkiye}}
\affiliation[3]{organization={Center of Excellence in Data Analytics}, addressline={Sabanci University}, city={Istanbul}, postcode={34956}, country={Türkiye}}
\affiliation[4]{organization={Department of Computer Science}, addressline={University of California, Santa Barbara}, city={Santa Barbara}, postcode={93106}, country={USA}}
\affiliation[5]{organization={Department of Computer Science}, addressline={ETH Zürich}, city={Zürich}, postcode={8092}, country={Switzerland}}
\affiliation[6]{organization={Department of Computer Science}, addressline={Université de Bordeaux}, city={Talence}, postcode={33405}, country={France}}
\affiliation[7]{organization={Department of Computer Science}, addressline={Columbia University}, city={New York}, postcode={10027}, country={USA}}

\begin{abstract}
Regular exercise is widely recognized as a cornerstone of health, yet sustaining consistent exercise habits remains challenging. Understanding the factors that influence the formation of these habits is crucial for developing effective interventions. This study utilizes data from Mars Athletic Club, Türkiye's largest sports chain, to investigate the dynamics of gym attendance and habit formation. The general problem addressed by this study is identifying the critical periods and factors that contribute to the successful establishment of consistent exercise routines among gym-goers. We show that specific periods of attendance are most crucial for habit formation. By developing a survival metric based on gym attendance patterns, we pinpoint these key phases and segment members into distinct clusters based on their visit patterns. Our analysis reveals significant differences in how various subgroups respond to interventions, such as group classes, personal trainer sessions, and visiting different clubs. Using causal inference analysis, we demonstrate that personalized guidance and social dynamics are key drivers of sustained long-term engagement. By systematically examining these variables and considering the specific characteristics of different clusters, our research highlights the importance of a tailored, multi-dimensional approach to promoting exercise habits, which integrates social dynamics, personalized guidance, and strategic interventions to sustain long-term engagement.
\end{abstract}

% You can include a graphical abstract here if required by the journal.

\begin{highlights}
\item We define a novel survival metric to measure the effect of gym attendance on habit formation. 
\item We identify critical 6-week and 17-week milestones for forming exercise habits. 
\item Gym members are clustered into 5 distinct behavioral groups using NMF. 
\item Personalized training and social dynamics are key drivers of long-term engagement. 
\item Tailored, multi-dimensional interventions are crucial for different member groups. 
\end{highlights}


\end{frontmatter}

\section{Introduction}\label{sec1}
Developing consistent exercise habits is one of today’s most pressing public health challenges, contributing to widespread inactivity and preventable chronic disease. Regular physical activity confers well‐established health benefits—improved cardiovascular and metabolic functioning, enhanced mental well‐being, and reduced risk of noncommunicable diseases—yet fewer than 25\% of adults worldwide meet recommended activity levels, and nearly 9\% of premature deaths are linked to sedentary lifestyles \citep{WHO2018,lee2012effect}. Despite widespread awareness of these benefits, many people struggle to translate their intentions into enduring exercise habits. One key reason is that starting an exercise routine requires consistent motivation, while maintaining that routine relies more on automatic processes. With repeated performance in stable environments, situational cues can trigger the behavior with little conscious effort. This phenomenon is known as cue-triggered automaticity \citep{Wood2016,gardner}.

Early work on habit formation showed that simple health behaviors (e.g.\ drinking water) reach plateaued automaticity in around 66 days of daily repetition, but more complex activities (like a full workout) take longer and vary widely across individuals \citep{lally}. In the exercise domain, longitudinal tracking of new gym members revealed that attending roughly four sessions per week for at least six weeks is the minimum “dosage” needed to establish a self‐sustaining gym habit—after which attendance becomes less dependent on conscious planning and more driven by contextual triggers \citep{kaushal2015}. However, mere repetition alone is not enough: positive affect experienced during workouts significantly boosts habit strength, whereas monotonous or aversive sessions fail to produce lasting automaticity \citep{weyland}. Conversely, individuals who abandon fitness trackers often cite lower internalized exercise motives and guilt about unmet goals, underscoring the need to cultivate intrinsic rather than purely extrinsic drivers during the habit-building phase \citep{nuss}. In the same vein, fitness-app users who report stronger “exercise identity” are substantially more active, suggesting that digital tools foster persistence by reinforcing self-concept as an exerciser \citep{barkley}. Recent machine‐learning analyses of millions of gym visits confirm that as exercise is consistently repeated in the same environment, individuals become less context‐sensitive—that is, their behavior shifts from deliberative goal pursuit toward habit‐driven action \citep{buyalskaya}. Recent evidence from wearable-technology users also shows that gamified tracking, instructional feedback, and progress sharing can accelerate the transition from intention to routine by strengthening self-determined motivation for physical activity \citep{gist}

Translating these insights into real‐world interventions requires moving beyond isolated strategies to comprehensive, multi‐level models. Socioecological frameworks posit that behavior is shaped simultaneously by individual traits, social networks, physical environments, organizational settings, and policy contexts \citep{sallis}. In practice, effective programs combine individual‐level techniques (e.g.\ skill training, self‐monitoring) with supportive social structures (group classes, workout partners), environmental cues (gym access, prompts), and even institutional policies (workplace wellness incentives) to create conditions under which exercise habits can form and endure \citep{sallis2018}. Moreover, talking about workouts both one-to-one with close contacts and in wider network broadcasts exerts distinct, additive influences on physical-activity levels, underscoring the need for interventions that enable both interpersonal and group communication \citep{limeng}. In online fitness communities, simply having a larger core network of active peers—and comparing one’s progress with fitter friends—predicts higher physical-activity levels, illustrating the motivational leverage of social ties \citep{huangg}. The COM‐B system further organizes this complexity by identifying three core determinants of behavior (capability, opportunity, and motivation) and mapping them onto intervention functions such as education, environmental restructuring, and enablement \citep{michie}. Integrated applications of socioecological and COM‐B models, for instance in office‐based physical activity initiatives, demonstrate how modifying workspace design, social norms, and individual skill can jointly enhance both the contextual cues and motivational drivers necessary for habit formation \citep{kasteren}.

Despite these theoretical advances, empirical studies rarely combine habit‐formation principles, multi‐level intervention design, and rigorous longitudinal analysis in authentic gym contexts. Most research focuses either on short‐term habit psychology under controlled conditions or on community‐level interventions without explicitly measuring automaticity. Meanwhile, survival‐analysis techniques offer powerful tools to model “time‐to‐habit” and persistence: by treating the establishment of a consistent workout pattern as a time‐to‐event outcome, one can quantify the hazard of relapse (dropout) and identify when habits are most vulnerable \citep{harris19,galvim}. A recent meta‐analysis of health‐behavior habit formation confirms that while 2–3 months of consistent action is typically required to reach plateaued habit strength, individual trajectories vary greatly and are influenced by factors such as routine timing, enjoyment, and environmental cues \citep{singh}.

Unlike laboratory habit studies that track reaction times or self‐report automaticity scales \citep{gardner, gardner2}, our large‐scale gym data lack those direct habit measures, so we use a survival‐based proxy for persistence instead. To our knowledge, no real‐world study to date has (a) segmented exercisers by temporal habit patterns, (b) mapped comprehensive, multi-level interventions onto those patterns, and (c) used time‐to‐habit survival models to quantify persistence. Habit strength is, at heart, persistence despite fluctuating motivation. Survival analysis models this directly by linking dropout risk to early cue–behavior repetition, turning habit formation into a time-to-event problem.

In this study, we bridge these gaps by using a large, real‐world dataset from Türkiye’s largest gym chain. We \textbf{(1)} define and apply a survival‐analysis–based metric to capture the persistence of gym attendance over time, \textbf{(2)} segment members into behavioral clusters based on their temporal visit patterns, and \textbf{(3)} employ causal‐inference methods to estimate the effects of diverse interventions—group classes, personal training, facility variety, and social invitations—while adjusting for demographic and contextual confounders. By integrating habit‐formation theory, multi-level intervention frameworks, and survival modeling, our work illuminates how and when exercise habits crystallize in a naturalistic fitness setting, and offers practical guidance for designing layered interventions that transform occasional gym‐goers into steady exercisers.

\section{Methods}\label{sec2}

\subsection{Dataset}\label{subsec:data-set}
Throughout this project, we analysed anonymized data provided by Mars Athletic Club. Mars Athletic Club (MAC) is the biggest gym chain in Türkiye, and has over 100 clubs in 13 provinces in the country, with most being in Istanbul, the most populated province. Founded in 2007, the number of locations has been growing steadily over the years.

Due to the COVID-19 pandemic, many countries implemented drastic measures. Türkiye was one of the countries that implemented lockdowns in 2020 and 2021, meaning that people experienced lockdowns and were unable to exercise indoors. These measures affected a lot of businesses, one of which was the fitness industry. As people were not allowed outside for extended durations, their gym memberships became obsolete during the pandemic.

Given the exceptional circumstances surrounding the data, we restricted our analysis to data from 2022 through the end of 2023. Following the insights of project partners at MAC, we concluded that it would be better if we considered only the first paid contract for all customers. This was done to account for users who may have been on subsequent contracts during this time frame, and this would mean that they would already have had a chance to achieve some habit formation before, which would give them an advantage over other members in forming habits. Additionally, we agreed to limit the contract type to annual, instead of having different contract types such as 6-month or monthly contracts. This would give every user an equal opportunity to either form habits within a reasonable time frame or fail to do so. Furthermore, we also filtered out non-paid contracts because these were usually allocated to employees. Crucially, all data used in this study were fully anonymized to safeguard personal information in compliance with Türkiye's Personal Data Protection Law (KVKK) \citep{kvkk2024}. This anonymization ensures the privacy of MAC's members while allowing for a comprehensive analysis of gym usage trends and behaviors without compromising individual confidentiality. No identifiable information was available to the research team at any stage of the analysis.

The variables used to analyse people’s habitual differences are listed below.

\begin{itemize}
    \item \textbf{Main club}: The club that the contract was signed at.
    
    \item \textbf{Membership Category}: Membership plan associated with the customer, which varies by club type and package.
    
    \item \textbf{Number of group classes attended}: Total number of group class participations by a customer. These classes may vary from club to club, but they are unified through various categories such as GFX, core, and cardio.
    
    \item \textbf{Number of sessions with a personal trainer}: The number of times that a customer worked with a personal trainer. These workouts are sold as sets of 10, and the customer may choose to train with any trainer they wish.
    
    \item \textbf{Number of different clubs visited}: As some contract types allowed for visiting multiple clubs, we counted the number of clubs visited during their contract lifetime.
    
    \item \textbf{Number of different group classes attended}: The number of distinct group class categories (e.g., GFX, core, cardio) attended by a customer.
    
    \item \textbf{Number of invitation credits used}: Total number of credits that the customer has used to invite their friends for a single gym session. Different numbers of credits are issued for different membership categories every month.
    
    \item \textbf{Previous experience level}: Self-reported prior experience level of the customer with fitness from MAC+ mobile application. Levels of 0 (beginner) through 3 (experienced). Available for around 40.2\% of the customers.
    
    \item \textbf{Form Level}: Self-reported current form level of the customer from MAC+ mobile application. For males, it is measured as the number of push-ups they can do in a minute, and for females, it is the number of bodyweight squats they can do in a minute. Levels of 0 (unfit) through 2 (very fit). Available for approximately 33.6\% of the customers.
    
    \item \textbf{Estimated Frequency of Exercise}: Self-reported estimated frequency of exercise from MAC+ mobile application. Levels of 0 (no exercise) through 2 (regular exercise). Available for more than 40\% of the customers.
    
    \item \textbf{Body Mass Index (BMI)}: Calculated from the self-reported weight and height from MAC+ mobile application. Available for all customers in our study.
\end{itemize}

\subsection{Clustering Customers}\label{subsec:cls-method}
We hypothesize that users may differ in their exercise habits, and these differences are often attributed to age and gender. However, external factors like exercise environment and daily routines also determine when people find opportunities for these activities. We used Non-negative Matrix Factorization (NMF) to identify distinct behavioral groups when studying daily and weekly gym visits. Different groups may have varying tendencies and independent habit formations, which we could incorporate to our analyses. The motivation for using NMF is its inherent ability to cluster data and its ease of interpretation compared to other clustering methods \citep{ding2005}. Furthermore, with this method, one can easily infer the cluster membership of input data. The NMF algorithm approximates the factorization of the data matrix $\mathbf{V}$ onto two non-negative matrices, $\mathbf{W}$ and $\mathbf{H}$, as shown in Equation \ref{eq:nmf}.

\begin{equation}
\mathbf{V} \approx \mathbf{W}\mathbf{H}
\label{eq:nmf}
\end{equation}where $\mathbf{V} \in \mathbb{R}_+^{m \times n}$ is the matrix with rows corresponding to features and columns to customers (observations), $\mathbf{W} \in \mathbb{R}_+^{m \times k}$ is the feature (basis) matrix, and $\mathbf{H} \in \mathbb{R}_+^{k \times n}$ is the coefficient matrix \citep{Kim2008SparseNM} with $k$ as a pre-specified parameter, denoting the number of components to represent the data matrix. Using the coefficient matrix $\mathbf{H}$, we assign data point $v_j$ to cluster $k$ if $\mathbf{H}_{kj}$ is the maximum across all components. To cluster the customers using NMF, we vectorized all the visits of each customer during their first six weeks by grouping them by the day of week and hour of day. This transforms each customer's visit history into a vector where each element corresponds to a specific day-of-week and hour-of-day time slot. The vectorization process is explained in detail in Appendix \hyperref[appendix:B]{B}.

As the first step, we conducted customer clustering on a single club that was representative of the whole customer base. Accordingly, we analysed one of the most popular MACFit branches in Istanbul, which had around 4,500 members at the time of the analysis. The company suggested this club, as they knew it resembled the general customer population. We experimented with a few values for $K \in \{3, \dots, 8\}$ to select the optimal number of clusters, and after $K = 5$, we observed diminishing returns in terms of reconstruction error. Additionally, larger values of $K$ produced clusters that were less interpretable. We thus settled on 5 clusters. Because the five-cluster solution separated the visiting patterns well, we applied NMF to the full customer base. Our results showed that 5 clusters were consistent throughout our whole customer base, thereby confirming the choice of $K$. The visit patterns of the customers are presented in Fig\ref{fig:clustering}(a).

Initially, customers were categorised strictly, with each customer assigned to exactly one cluster. However, later on, since we would be using propensity score matching (PSM), having the probabilities of belonging to each cluster would provide significant benefits for the matching algorithm. Thus, we utilized the output of the NMF algorithm to get the probabilities. The coefficient matrix $\mathbf{H}$ can be used either for hard assignment of customers to clusters or for calculating membership probabilities. We used a softmax function across the $K$ components for each customer to calculate the probability of belonging to the clusters as follows.

\begin{align*}
    P(z_j = k) = \frac{\exp(\mathbf{H}_{kj})}{\sum_{i=1}^K \exp(\mathbf{H}_{ij})}
\end{align*}This normalization ensures a valid probability distribution, with values between 0 and 1 that sum to 1 across all $K$ clusters for each customer. The resulting vectors capture how strongly each of the $K$ visit patterns contribute to each customer's visit profile. The cumulative distribution functions for the probabilities of weeks 6 and 17 are presented in Figures \ref{fig:clustering}(c,d).

\subsection{Cluster-level Deviations for Demographic Groups}\label{subsec:cls-dev-dem}

In each behavioral cluster, user demographics may differ, with certain age and gender groups more likely to be observed in some clusters. To distinguish the prior probability of observing certain user demographics ($D_i$) from their preferential occurrences, we calculated the likelihood ratios between conditional probability of being in a given cluster ($C_i$) following Eq.~\ref{eq:logratio}

\begin{equation}
\text{Likelihood ratio for demographic $D_i$ in cluster $C_j$} =
\frac{P(D_i \mid C_j)}{P(D_i)}
\label{eq:logratio}
\end{equation}This value will be 1 if cluster assignment has no effect on observing certain demographic groups. A deviation from 1 indicates whether a group is more ($>1$) or less ($<1$) likely to be observed. We obtained the deviations by subtracting 1 from this value.

\subsection{Defining a Survival Metric for Habit Formation}\label{subsec:sur-method}
Understanding the underlying mechanisms of habit formation is key to developing targeted interventions that motivate individuals to attend the gym consistently. We developed the survival metric to measure and analyze the persistence of gym habits. Our approach to this analysis was inspired by the work of Harris and Kessler (2019), who explored habit formation and activity persistence in the context of gym equipment usage. Their study demonstrated that frequent early activity leads to more persistent exercise behavior, suggesting that interventions aiming for behavioral change need to involve higher frequency or longer duration activities to be effective \citep{harris19}. This metric was defined as the number of consecutive weeks in which a member attended the gym at least once. We added a tolerance of one week of absence called gap week, acknowledging that occasional interruptions (e.g., travel or illness) should not be taken as a disruption of habit formation. The decision to include this tolerance was based on an analysis of intermediate periods between consecutive gym attendance periods, which can be found in Appendix \hyperref[appendix:C]{C}, which showed that a one-week gap was most commonly associated with sustained attendance.

However, due to the inclusion of this tolerance period, the number of weeks it took for individuals to reach each survival milestone (i.e., 6, 7, \dots ,17 weeks) varied across users. As a result, when creating attendance milestones, it was imperative to assess users' habit formation based on the number of visits made during a fixed period rather than the survival longevity. This approach also provided a fixed period for later analysis of users' behaviors and allowed us to investigate the impact of early actions on long-term persistence.

To further assess and track habit formation, we established specific attendance milestones corresponding to critical survival durations. These milestones mark a member's progress in their gym attendance journey. Each milestone is defined by a critical visit count that members must achieve within the specified period. For example, to reach the $\text{6}^\text{th}$ week milestone, a member must attend the gym at least 9 times within the first six weeks of their membership. Figure \ref{fig:critical_visits}(d) shows an almost linear trend in the critical number of visits required to reach milestones. We fitted a linear model \( y = 2.01x - 5.35 \), with \( \text{slope} \approx 2.01 \), indicating that members need to visit the gym roughly twice per week to meet their milestones. This suggests that attending the gym at least twice a week is important for maintaining consistent habits. These milestones were introduced to provide a structured approach for evaluating habit formation over time. Users may still attend the gym for extended periods, but these milestones allow us to separate users with the same length of membership into two categories. By focusing on these milestones, we aimed to simplify the tracking of member progress and to identify points at which interventions might be most effective. 

\begin{figure}[H]
\centering
\includegraphics[width=1\textwidth]{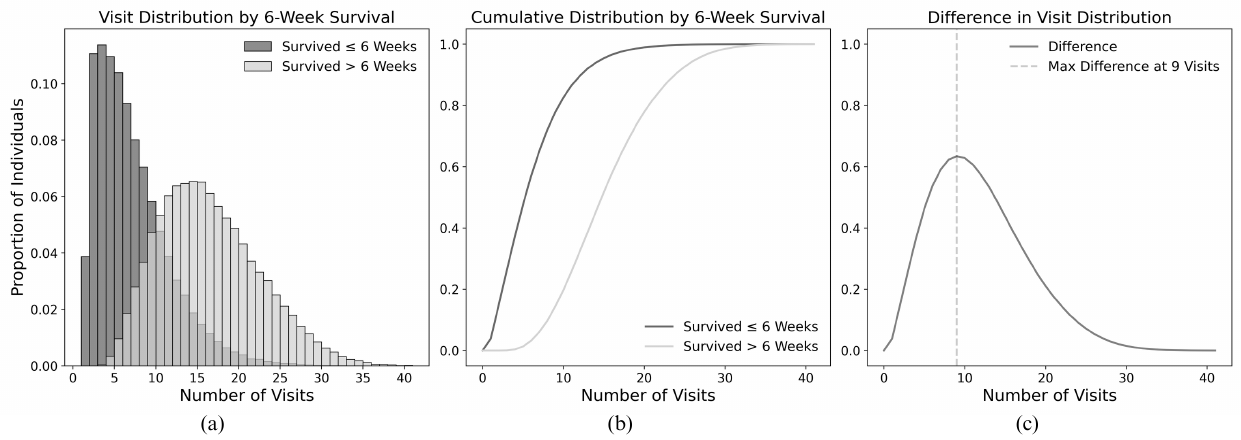}
\caption{\textbf{Estimating critical number of visits}. We compared the visit distributions of the members who survived less than or equal to 6 weeks and survived more than 6 weeks (a) by looking at their cumulative distributions (b) to determine the critical visit threshold that best separates the two groups (c).}
\label{fig:survival}
\end{figure}

To determine the critical minimum number of visits required to reach a 6-week survival milestone, we compared the distribution of the number of visits made in 6 weeks by the members who survived 6 weeks or less and who survived more than 6 weeks in Figure \ref{fig:survival}(a). We analyzed the cumulative distribution of visits for both groups, visualized in Figure \ref{fig:survival}(b), examined the difference between the cumulative distribution functions (CDFs), and determined the number of visits that maximized the difference between the two groups, as shown in Figure \ref{fig:survival}(c). The critical number of visits was nine within the first six weeks, as this threshold maximized the difference between those who maintained their gym habits and those who did not. This approach allowed us to identify the number of visits that best demonstrated sustained engagement and habit formation during the critical early weeks of membership. Critical visit counts for each week, weeks 6 through 52, are included in Figure \ref{fig:critical_visits}(d).

\subsection{Causal Inference Framework}\label{subsec:causal-methods}
The fundamental objective of our investigation is to understand the causal effects of diverse gym-related interventions on the establishment of consistent exercise habits. While traditional correlation-based methodologies provide valuable insights, they remain inherently limited in their capacity to distinguish genuine causal relationships from mere statistical associations. Initially, we employed a supervised learning approach, constructing a logistic regression model to predict six-week survival (defined as whether a member attained the critical visit threshold) as a function of early interventions and relevant covariates (including age, gender, BMI, behavioral cluster, membership details, and self-reported form/experience/frequency metrics). This preliminary model revealed statistically significant positive associations; increased group lesson attendance, more frequent personal trainer sessions, and greater club variability were all correlated with enhanced odds of survival. However, this approach could not effectively disentangle whether these relationships were causally meaningful or merely reflections of confounding variables (for instance, intrinsically motivated individuals might simultaneously select more interventions and demonstrate greater persistence). Furthermore, standard logistic regression treats all variables with symmetrical importance and lacks an explicit mechanism to encode structural assumptions regarding which factors might confound specific treatment–outcome relationships. While valuable for predictive purposes, such supervised methods fundamentally lack the formal identification and rigorous adjustment strategies essential for robust causal interpretation.

To address these methodological limitations, we utilized a causal inference framework based on clear graphical assumptions and techniques for adjusting propensity scores. This approach allows us to estimate the effects of interventions under counterfactual conditions. It effectively measures outcomes as if participants had been randomly assigned to different treatment levels, thereby isolating the true causal impact from misleading correlational artifacts.

We defined our treatment variables $T \in \mathcal{T}$ as systematically binned levels of five distinct gym interventions measured over the initial six-week period: group lessons, personal-trainer sessions, variety of lesson types, diversity of club visits, and invitation credit usage. Additionally, we incorporated three self-reported variables (form level, experience level, and estimated visit frequency) as supplementary treatment indicators. The binary outcome variable $Y \in \{0,1\}$ indicates whether a member successfully achieved the six-week critical-visit milestone. Potential confounding variables $\mathbf{X} \in \mathcal{X}$ encompass demographic factors (age, gender, BMI), membership characteristics (start date, primary club affiliation, membership category), and initial behavioral cluster assignment. These covariates were selected based on their theoretical potential to simultaneously influence both the probability of receiving specific treatments and the likelihood of establishing consistent attendance patterns.

We formalized our causal assumptions through a directed acyclic graph (DAG) $\mathcal{G} = (\mathcal{V}, \mathcal{E})$ where $\mathcal{V}$ represents nodes and $\mathcal{E}$ directed edges. In this graph, edges connect confounders $X$ to both treatment $T$ and outcome $Y$, with additional edges directly linking $T$ to $Y$. Under this causal structure, the average treatment effect (ATE) of setting $T=t$ is identified through the backdoor adjustment formula:
\begin{align}
\mathbb{E}[Y \mid \mathrm{do}(T=t)] &= \sum_{x \in \mathcal{X}} \mathbb{E}[Y \mid T=t, \mathbf{X}=x]\,P(\mathbf{X}=x)\\
&= \mathbb{E}_{\mathbf{X}}[\mathbb{E}[Y \mid T=t, \mathbf{X}]]
\end{align}
where $\mathrm{do}(T=t)$ represents Pearl's do-calculus operator indicating intervention rather than passive observation. We verified that conditioning on $\mathbf{X}$ effectively blocks all non-causal backdoor paths between $T$ and $Y$, thereby satisfying the backdoor criterion for causal identification.

To implement the backdoor adjustment efficiently, we estimate propensity scores $e(\mathbf{x})=P(T=1\mid \mathbf{X}=\mathbf{x})$ via logistic regression and execute one-to-one nearest-neighbor matching without replacement, employing the following estimator:
\begin{align}
\hat{e}(\mathbf{x}) &= \frac{\exp(\mathbf{x}^T\boldsymbol{\beta})}{1 + \exp(\mathbf{x}^T\boldsymbol{\beta})}
\end{align}
where $\boldsymbol{\beta}$ represents the vector of regression coefficients. This matching procedure effectively balances the joint distribution of $\mathbf{X}$ between treated and control groups, as confirmed by standardized mean differences consistently below the conventional threshold of 0.1 across all covariates. The resulting matched sample approximates the conditions of a randomized experiment, enabling unbiased estimation of the ATE through:
\begin{align}
\widehat{\mathrm{ATE}} &= \frac{1}{N}\sum_{i=1}^N \bigl(Y_i^{(T=1)} - Y_i^{(T=0)}\bigr)\\
&= \frac{1}{|\mathcal{M}|}\sum_{i \in \mathcal{M}} \bigl(Y_i - Y_{m(i)}\bigr)
\end{align}
where $Y_i^{(t)}$ denotes the potential outcome for unit $i$ under treatment level $t$, $\mathcal{M}$ represents the set of treated units that were successfully matched, and $m(i)$ denotes the matched control unit for treated unit $i$.

To establish well-defined treatment conditions, continuous and count-based intervention variables are systematically categorized into four distinct bins: \texttt{None} (no intervention), \texttt{Low} (intervention level $\leq$ 33rd percentile), \texttt{Moderate} (intervention level between 33rd and 66th percentiles), and \texttt{High} (intervention level $>$ 66th percentile). Self-reported ordinal variables with $n$ levels are binarized using $n-1$ threshold indicators (e.g., "form level $\geq 1$" versus "0," followed by "form level $= 2$" versus "$\leq 1$"), thereby preserving the inherent ordering while ensuring adequately balanced sample sizes across treatment conditions.

We used the DoWhy library \citep{dowhy} to implement our causal inference pipeline through a systematic workflow:
\begin{enumerate}
    \item Specify the dataset triplet $(Y, T, \mathbf{X})$ and formalize the causal structure via DAG notation.
    \item Automatically verify the applicability of backdoor adjustment through d-separation tests.
    \item Estimate propensity models and execute matching procedures with appropriate diagnostics.
    \item Calculate point estimates of ATEs and construct nonparametric bootstrap confidence intervals.
\end{enumerate}

To ensure the reliability of our causal estimates against potential methodological artifacts or unobserved confounding, we conducted sensitivity analysis as well. We introduce randomly generated variables as pseudo-treatments, confirming that no significant causal effects are detected (yielding $p$-values $\approx 1.0$ as expected under the null hypothesis).

To capture potential heterogeneity in treatment effects across behavioral patterns, we also independently replicated the propensity score matching (PSM) procedure within each of the five NMF-derived behavioral clusters. In these stratified models, we exclude the cluster label from the confounder set $\mathbf{X}$ to prevent collinearity issues, thereby enabling unbiased estimation of cluster-specific treatment effects through:
\begin{align}
\widehat{\mathrm{ATE}}_c &= \frac{1}{|\mathcal{M}_c|}\sum_{i \in \mathcal{M}_c} \bigl(Y_i - Y_{m_c(i)}\bigr)
\end{align}
where $\mathcal{M}_c$ represents the matched treated units within cluster $c$, and $m_c(i)$ denotes the matched control unit for treated unit $i$ within that cluster.

\begin{table}[ht]
\caption{Covariates used for confounding adjustment in PSM for each intervention.}\label{tab:interventions_used}
\resizebox{\columnwidth}{!}{%
\begin{tabular}{|l|cccccccc|}
\hline
& \multicolumn{3}{c|}{\textbf{Demographic}} & \multicolumn{3}{c|}{\textbf{Membership}} & \multicolumn{2}{c|}{\textbf{Experience}} \\ \hline
\textbf{Intervention} & Age & Gender & BMI & Start & Club & Category & Form & Cluster \\ \hline
Group Lessons & \checkmark & \checkmark & \checkmark & \checkmark & \checkmark & \checkmark & \checkmark & \checkmark \\
Personal Trainer Sessions & \checkmark & \checkmark & \checkmark & \checkmark & \checkmark & \checkmark & \checkmark & \checkmark \\
Invitation Credits & \checkmark & \checkmark & \checkmark & \checkmark & \checkmark & \checkmark & \checkmark & \checkmark \\
Different Club Visits & \checkmark & \checkmark & \checkmark & \checkmark & \checkmark & \checkmark & \checkmark & \checkmark \\
Different Group Lessons & \checkmark & \checkmark & \checkmark & \checkmark & \checkmark & \checkmark & \checkmark & \checkmark \\
\hline
\multicolumn{9}{|l|}{\textbf{Self-Reported Variables}} \\
Form Level & \checkmark & \checkmark & \checkmark & \checkmark & \checkmark & \checkmark & & \checkmark \\
Experience Level & \checkmark & \checkmark & \checkmark & \checkmark & \checkmark & \checkmark & & \checkmark \\
Estimated Visit Frequency & \checkmark & \checkmark & \checkmark & \checkmark & \checkmark & \checkmark & & \checkmark \\
\hline
\end{tabular}%
}
\end{table}

By integrating preliminary associative modeling with a specified causal inference framework; we derive methodologically sound and reproducible estimates of how early gym interventions \emph{causally} influence the establishment and maintenance of consistent exercise habits.

\section{Results}\label{sec3}

\subsection{Clustering Based on Visits}\label{subsec:cls-res}

\begin{figure}[H]
\centering
\includegraphics[width=\textwidth]{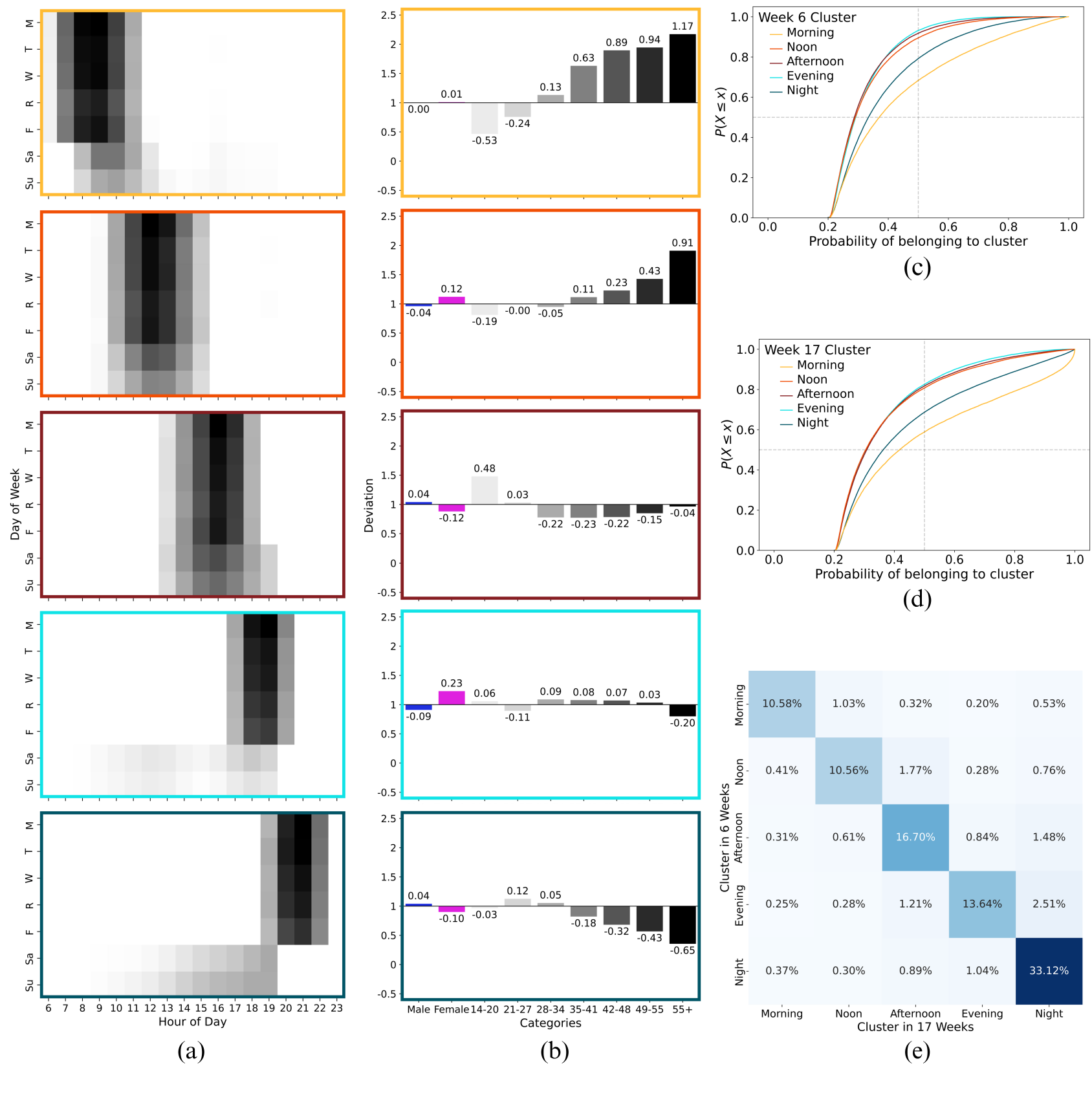}
\caption{\textbf{Behavioral clusters and preferences by demographics}. NMF reveals 5 distinct exercise groups. These groups differ mainly by their preference to exercise during the day (a). Some of these clusters are more preferred by some gender and age groups (b). Since the cluster assignments were determined by user activities in the first 6 weeks, we measured the confidence of the NMF model for each cluster (c,d). Finally, we check the consistency of assignment when data for the first 6 and 17 weeks are considered (e).}
\label{fig:clustering}
\end{figure}

Clustering analysis revealed five distinct clusters across all clubs, as shown in Figure \ref{fig:clustering}(a). We identified five clusters corresponding to distinct daily time slots: morning (08:00–09:00), noon (12:00–13:00), afternoon (15:00–17:00), evening (18:00–19:00), and night (20:00–21:00). The morning and night clusters also exhibited distinct weekend patterns, which reflects the reduced weekend opening hours (08:00–20:00) compared to weekdays (06:00–23:00).

These behavioral groups were labeled according to their peak visit hours. The densities of the visits presented in the heatmaps reflect the most preferred times to visit the gym for each cluster. For this analysis, clustering was performed using only the first six weeks of visit data for each customer; however, we later demonstrate that the assigned clusters remain largely consistent when compared to visits through the $\text{17}^\text{th}$ week.

We then defined cluster-level deviations as in Section \ref{subsec:cls-dev-dem}. With this definition, we wanted to see how the clusters deviated with respect to different variables we could group the customers, such as their gender and age groups. The deviation measures how much more or less likely it is to observe particular type of users within that cluster. For example, a deviation of 0.01 indicates that women are 1\% more likely to appear in the Morning cluster compared to the overall population, as presented in the first subfigure in Fig\ref{fig:clustering}(b). Conversely, individuals aged 14–20 are 53\% less likely to be found in the Morning cluster, as depicted in the same subfigure.

Younger people (ages 14-27) tend to make fewer visits in the morning and noon, as indicated by their negative deviation in the two upper subfigures in Fig\ref{fig:clustering}(b). Conversely, older individuals (ages 49+) tend to visit more during the earlier parts of the day, as their morning deviations are positive while the rest are negative. This may be attributed to a preference for visiting before work to free up time afterward. The larger magnitude of deviations partly reflects the smaller sample size of older members. Women are more prominent in the noon and evening clusters, while men, as they make up the majority of people, do not deviate too much, but visit more often in the afternoon and at night. Most starkly, the people aged 14-20 make most of their visits in the afternoon. This likely reflects their school or university schedules, which permit afternoon gym attendance.

Figures \ref{fig:clustering}(c) and (d) show the cumulative distribution function for the probability of belonging to a cluster for each different cluster, signified by their unique color. These figures show that the NMF decomposition assigns morning and night clusters with higher confidence, whereas assignments for the intermediate clusters were less confident. This is also evident by the heatmaps produced in Fig\ref{fig:clustering}(a), as there is substantial overlap for the rest of the clusters. We hypothesize that individuals in the morning and night clusters may have more fixed schedules, which prevents them from visiting at other times. The rest of the clusters have more flexible schedules, allowing them to change their gym hours more frequently, thus making them harder to assign to a single cluster.

Figure \ref{fig:clustering}(e) presents the cluster transition matrix, where the diagonal values represent the percentage of the population that remained in the same cluster between weeks 6 and 17. Although the previous two figures showed that morning and night clusters were more confidently assigned, this figure demonstrates that cluster assignments are robust over time. The diagonal values, which account for more than 84\% of the population, indicate that members generally do not transition between different behavioral clusters. Additionally, most of the transitions are between clusters with adjacent time slots, which is to be expected. Thus, even though assignment is less confident for intermediate clusters, the observed stability suggests that underlying temporal preferences are persistent traits rather than random fluctuations. In sum, these clusters capture persistent, demographically patterned exercise routines. Their stability over time provides a robust foundation for subsequent causal inference and intervention analyses.

\subsection{Identifying Critical Milestones for Habit Formation}\label{sec:critical-visit}
\begin{figure}[H]
\centering
\includegraphics[width=\textwidth]{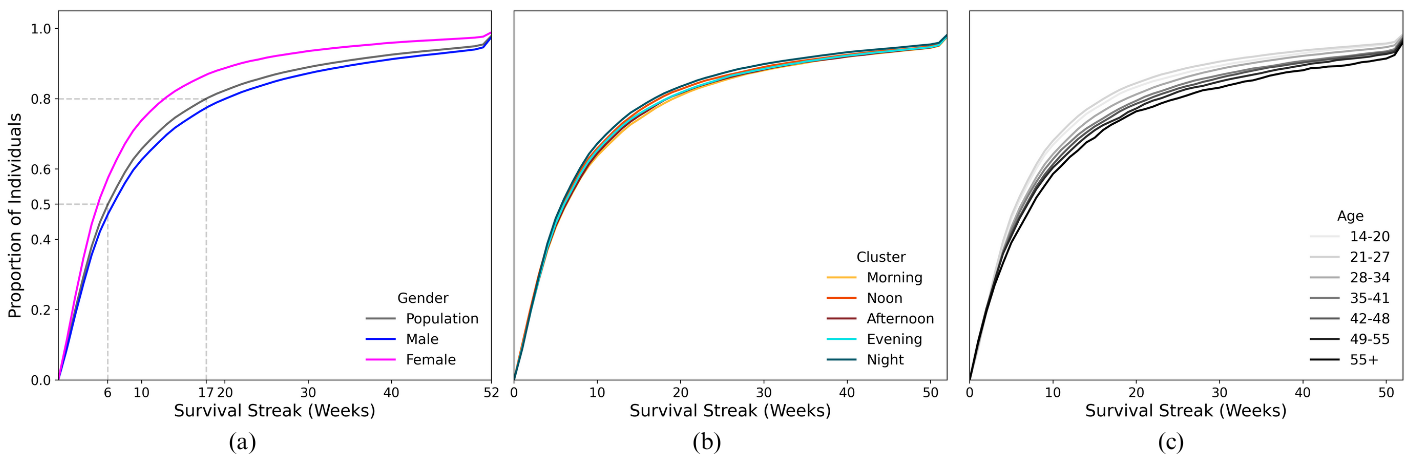}
\caption{\textbf{Cumulative distributions of the members' survival streaks by clusters and demographics}. We looked at the cumulative distributions of the length of the survival streaks for the whole population and genders (a), different behavioral clusters (b), and stratified groups by age (c).}
\label{fig:survival_cumulative}
\end{figure}

\begin{figure}[H]
\centering
\includegraphics[width=\textwidth]{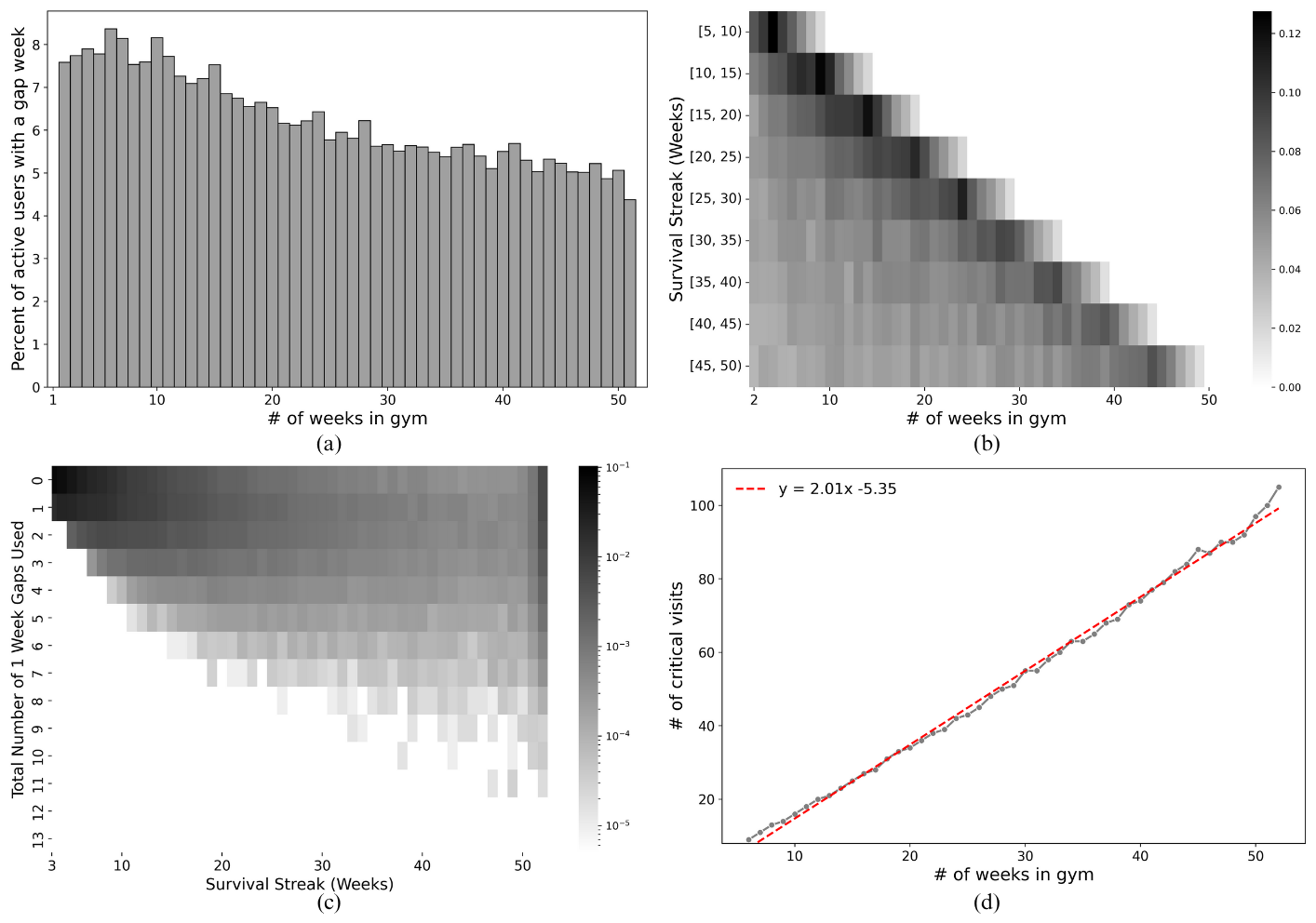}
\caption{\textbf{1 Week Allowance Analysis and Critical Visit to Reach Milestones}. Since members are allowed to take 1-week gaps during their 52-week journey, we count gaps used for each week (a). To see in which weeks the members take a 1-week gap, we divided them into subpopulations of different survival bins and analyzed the distribution of gaps (b). Members are allowed to use gaps more than once, and we checked the distribution of the total number of used gaps by the length of the survival streaks (c). Survival analysis results in critical visits for each week in the member's journey (d)}
\label{fig:critical_visits}
\end{figure}

Our analysis of the survival metric revealed clear patterns in gym attendance behavior and allowed us to identify essential milestones that indicate the progression of habit formation. As shown in Figure \ref{fig:survival_cumulative}(a), the cumulative distribution of members’ survival series shows a rapid increase in gym attendance in the first weeks. This rapid increase may reflect high initial motivation. To structure our interpretation, we aimed to define key milestones in this decline. Specifically, we identified the weeks at which 50\% and 80\% of users had ceased to maintain their attendance streaks. These thresholds were not selected based on any external theoretical rationale but served as pragmatic markers to summarize habit formation and dropout trends. Using this approach, we observed that approximately 50\% of members did not maintain their attendance streak at the 6-week point. This 6-week milestone emerged as a critical point in the habit formation process where members either solidify or begin to abandon their gym habit. When we extended our analysis to users who survived even longer, we observed a steady decline in the number of members maintaining their attendance streak. At the 17-week point, 80\% of the member population failed to maintain their streak, indicating an increasing difficulty in maintaining long-term attendance.
Furthermore, our analysis showed that to pass this critical 6-week milestone, a member typically needed to make at least nine visits, establishing an average frequency of roughly two visits per week as a key behavior for sustained attendance (Figure \ref{fig:critical_visits}(d)).

In Figure \ref{fig:survival_cumulative}(b), we examined the survival metric across different user clusters. Our analysis revealed no significant difference in habit persistence across these clusters. As shown in Figure \ref{fig:survival_cumulative}(a), when we analyzed the survival metric by gender, we observed a significant difference between male and female members. Male members showed longer survival streaks, indicating marginally better performance in gym attendance over time. Finally, Figure \ref{fig:survival_cumulative}(c) examines the survival metric across age groups. The analysis shows that older members tend to maintain their gym habits longer than younger ones. This comparison across different groups suggests that when people exercise during the day, it has less effect on habit formation than their demographics. We will further investigate these dimensions in our causal analysis.

Figure \ref{fig:critical_visits}(a) shows that while there is a slight decrease, the distribution of gap week usage remains relatively uniform overall. Although the general trend might seem to decrease, as seen in Figure \ref{fig:critical_visits}(b), the reason for the apparent higher gap usage in the initial weeks is due to the significant proportion of individuals with short survival streaks, as previously discussed, with 50\% having a survival value of less than 6 weeks. However, as illustrated in Figure \ref{fig:critical_visits}(b), for all survival subgroups, the rate of gap usage seems to increase as they approach the end of their survival streak, indicating a higher likelihood of churn. Figure \ref{fig:critical_visits}(c) reveals that most users do not use more than a few gaps. Moreover, users' reliance on gap weeks decreases as they achieve longer survival streaks. As the gradient shading suggests in Figure \ref{fig:critical_visits}(c), the likelihood of members using gap weeks decreases as survival streaks lengthen, which could imply that habit strength increases with consistent attendance.

\newpage
\subsection{Impact of Interventions on Habit Formation}
\begin{figure}[H]
\centering
\includegraphics[width=1.0\linewidth]{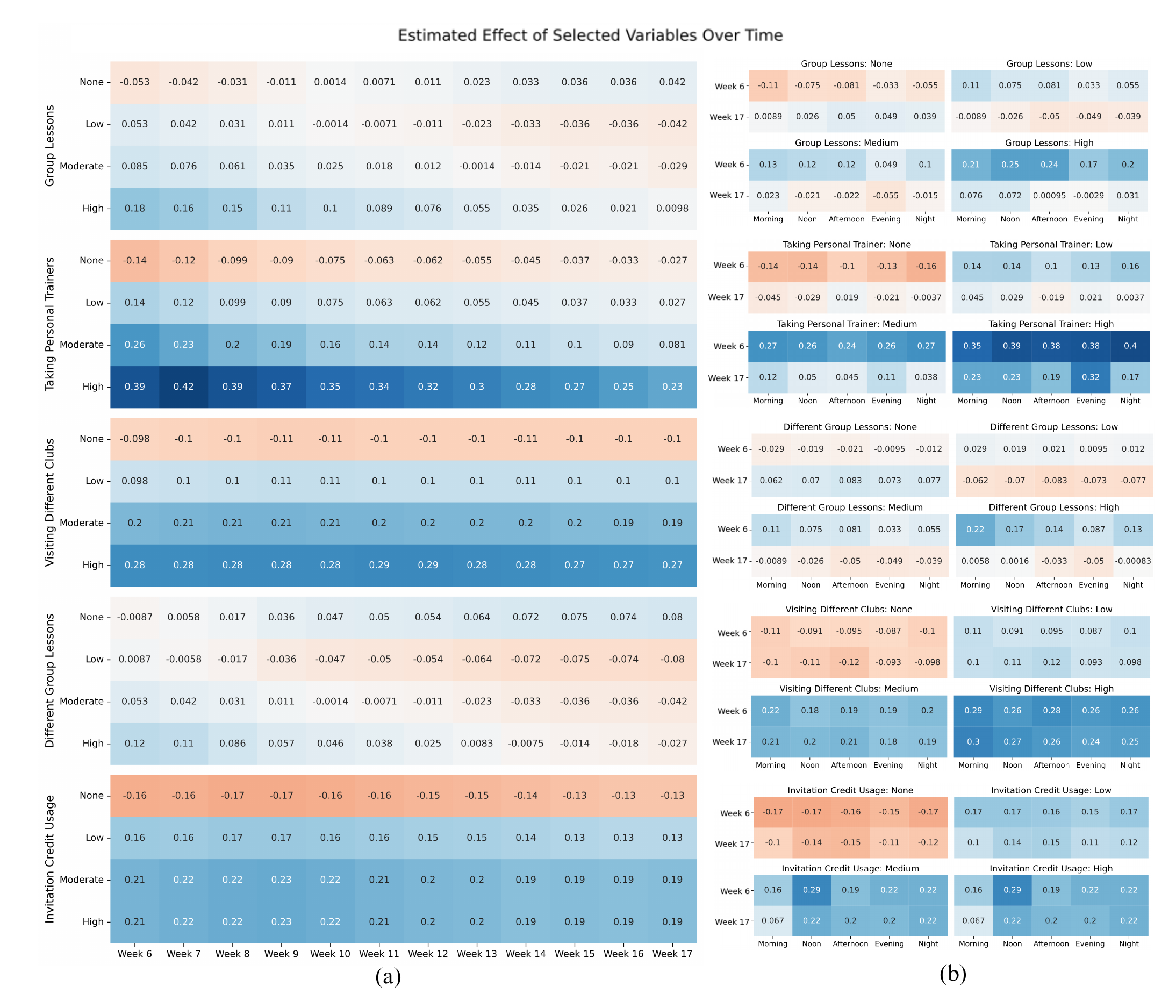}
\caption{\textbf{Estimated effects of interventions on habit formation}. The overall effects of various gym-related interventions on habit formation, shown from the $\text{6}^\text{th}$ to the $\text{17}^\text{th}$ week (a). The impact of these interventions across different clusters, specifically highlighting the effects observed at the $\text{6}^\text{th}$ and $\text{17}^\text{th}$ weeks (b).}
\label{fig:causal_inference}
\end{figure}

Figure \ref{fig:causal_inference}(a) shows the overall effects of various gym-related interventions on habit formation.

The effect of \textbf{attending group lessons} starts relatively high but decreases over time. Despite this decline, the impact remains substantial. Notably, members who attend group lessons at the highest level exhibit significantly greater positive effects on habit formation compared to those at lower levels. The communal aspect of group lessons likely provides both social support and accountability, which are critical in the early stages of habit formation. Additionally, \textbf{the number of different group classes attended} shows a similar pattern, with a positive effect on habit formation, though slightly less pronounced compared to the overall attendance at group lessons. This suggests that while variety in group class participation contributes to habit formation, the sheer frequency of participation plays a more dominant role.

\textbf{Personal trainer sessions} exhibit a strong positive impact on maintaining regular exercise habits. The coefficients for personal training are consistently higher than those for other interventions, indicating the significant role of personalized guidance and accountability in habit formation. Personal trainers offer customized workout plans and motivational support, which appear to be significant in helping members establish and maintain their exercise routines.

The effect of \textbf{visiting different clubs} is also substantial initially and remains one of the stronger effects among the interventions. This suggests that variety in gym environments significantly supports the development of exercise habits. The exposure to different facilities and equipment likely adds an element of novelty and excitement, preventing the monotony that can sometimes lead to dropout.

\textbf{Using invitation credits} shows a consistent positive effect on gym attendance over time. Unlike other interventions like taking personal trainer and group lessons whose impact may diminish as individuals establish their routines, the effect of inviting friends to the gym remains relatively stable. This suggests that social interactions and shared experiences continue to play an important role in motivating regular attendance, even as individuals settle into their exercise habits. The sustained impact of invitation credit usage highlights the importance of utilizing social connections to maintain engagement and support long-term habit formation.

The results of the cluster analysis also indicate that specific interventions have diverse impacts on distinct subgroups of gym members. The impact of these interventions across different clusters, specifically highlighting the effects observed at the $\text{6}^\text{th}$ and $\text{17}^\text{th}$ weeks shown in Fig\ref{fig:causal_inference}(b).

\textbf{Morning cluster} members show more positive responses to taking different group lessons and personal training sessions compared to other clusters. In this cluster, older adults are over-represented relative to the general gym-going population as presented in an earlier section. The variety offered by different group lessons plays a crucial role in maintaining their engagement, as it may introduce novelty and keeps the routine from becoming monotonous. These members benefit greatly from structured and personalized interventions that provide both variety and individualized support. The consistent early morning routine likely helps in establishing a disciplined approach to exercise, which is particularly important for older adults who may prioritize health and wellness.

\textbf{Noon cluster} members exhibit a significantly higher positive response to invitation credit usage and group lessons compared to other clusters. This group tends to be older adults, with a higher-than-average proportion of females. Since the noon cluster hours coincide with working hours in Türkiye, it is possible that noon cluster members have more flexible schedules. The flexible schedules of the noon cluster may allow them to view the gym as a social hub, where they can combine exercise with socializing, which is particularly appealing to older adult members. Additionally, these members show a sustained positive effect from taking personal trainer sessions, particularly by week 17. This indicates that personalized guidance and accountability provided by personal trainers continue to support consistent exercise habits over time, making it an important intervention for this group.

\textbf{Afternoon cluster} members are predominantly within the 14-20 age group, making younger gym-goers over-represented in this cluster compared to others. This demographic is often more experimental and seeks variety in their activities, which explains their strong positive response to visiting different clubs. Interestingly, the effectiveness of working with a personal trainer in this cluster is quite high initially but shows a strong decrease over time. This may indicate that younger members quickly benefit from the initial guidance and motivation provided by a personal trainer but may lose interest or feel more confident in continuing on their own after the early weeks. Additionally, group lessons have a consistently positive impact on this cluster, providing a structured environment that may help sustain their engagement by offering social interaction and variety.

\textbf{Evening cluster} members have the highest proportion of females, which influences their response to interventions. Unlike other clusters, this group shows a lower positive response to group lessons. Although the female presence is significantly higher, similar to the noon cluster, the cluster behavior differs significantly. The overall effect of group lessons on the evening cluster is negligible, with the impact being lower than the entire population (0.18), suggesting that group workouts may not be as appealing to them. However, this cluster shows the highest and most sustained positive effect from personal trainer sessions by the $\text{17}^\text{th}$ week, making it the cluster that best maintains its benefits from personalized training. This indicates that for evening exercisers, particularly in a female-represented cluster, personalized attention and tailored workouts play a crucial role in fostering long-term exercise commitment, especially during the evening hours after a day's work.

\textbf{Night cluster} benefits from a mix of interventions, including group lessons and personal training sessions. The late-night hours are typically suitable for those with unconventional schedules, such as students or shift workers. The fact that this group is predominantly younger, with older individuals being strongly underrepresented also supports this. Working with a personal trainer has the most significant effect on habit formation within this cluster during the first 6 weeks. However, this effect diminishes considerably by the $\text{17}^\text{th}$ week. This decline may suggest that younger members, who form the majority of this cluster, initially benefit from the structure and guidance provided by a personal trainer but may choose to transition to more independent workouts as they gain confidence and familiarity with their routines. The initial high impact of personal training likely helps them establish a solid exercise habit, but as their need for external motivation decreases, they may feel less inclined to continue using these services.

\begin{figure}[H]
\centering
\includegraphics[width=1.0\linewidth]{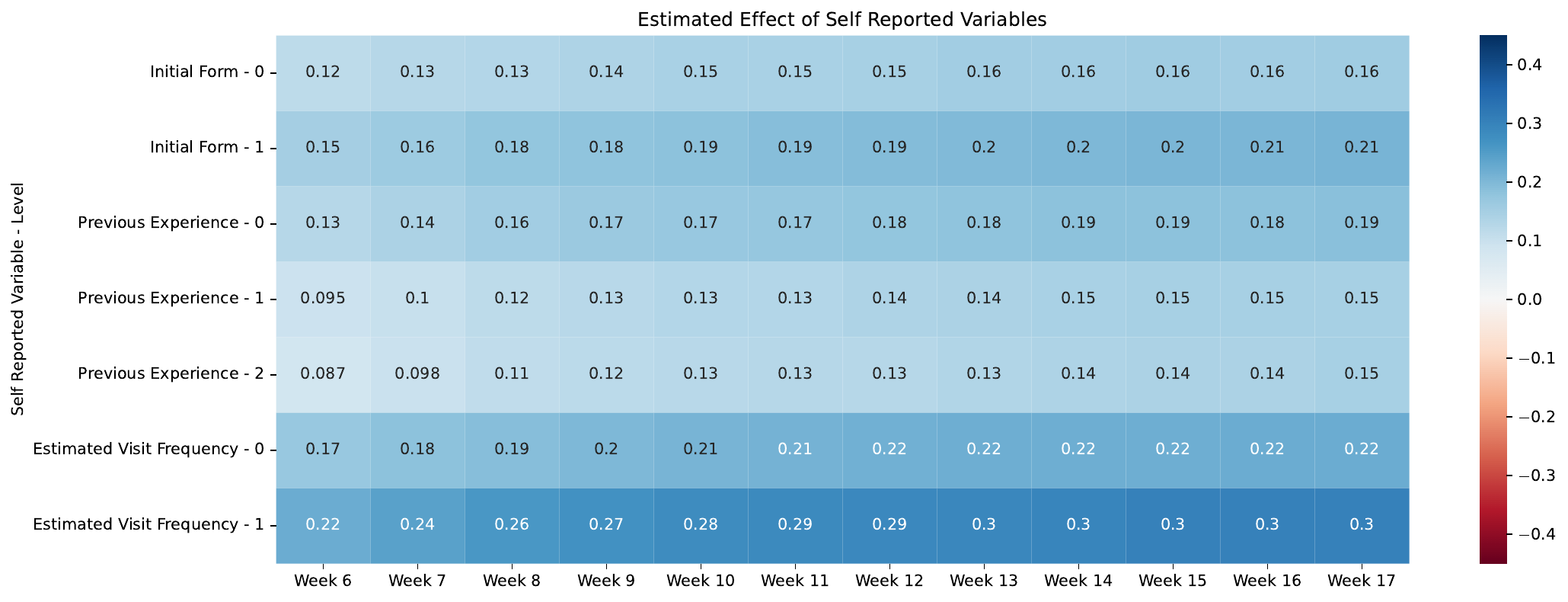}
\caption{\textbf{Estimated effects of self-reported variables on habit formation.} User-reported characteristics separated by their intensity for previous form, experience and visit frequencies.}
\label{fig:gtkb}
\end{figure}

\newpage
Figure \ref{fig:gtkb} presents the estimated effects of self-reported variables like form level, experience level, and estimated visit frequency over time. These self-reported variables provide insights into how individual characteristics influence the formation of consistent exercise habits, offering a different perspective compared to the direct intervention variables analyzed earlier. These variables are gathered from users through a survey on the mobile app at the start of their membership.

The effect of the \textbf{initial form level} on habit formation shows a gradual and consistent increase over time. Members who reported higher form levels at the start were more likely to develop and maintain consistent exercise habits. This contrasts with the impact of direct interventions like group lessons or personal training sessions, where the effect tends to peak early and may diminish over time. The steady rise in the effect of form level suggests that physical fitness provides a robust foundation for sustaining long-term exercise routines, potentially due to greater confidence and physical capability, which enable members to stick to their routines more effectively.

\textbf{Prior experience with fitness} exhibits a steadily growing effect on gym attendance consistency over time as well. Interestingly, we do not observe that same clear "more is better" trend among those claiming to have exercised previously: in fact, people with no experience end up slightly surpassing the other experience levels. This phenomenon may hints at a beginner’s boost, wherein newcomers may benefit from heightened early support or enthusiasm that temporarily exceeds the advantages one might expect from prior fitness background. Another possible explanation is that individuals who have never engaged in fitness before may find it easier to establish new habits compared to those who have started and stopped exercising in the past, as the latter group might face psychological barriers or carry over negative associations. Despite this, we still see consistent effects among members with previous fitness experience.

The initial \textbf{estimated visit frequency} is another self-reported variable that demonstrates a continuously increasing effect on habit formation. Members who were estimated to visit the gym more frequently at the outset were more likely to sustain their exercise routines. The positive trajectory of visit frequency’s effect suggests that regularity in the initial stages of gym attendance is critical for embedding exercise as a long-term habit. Unlike some interventions that provide an initial push, a high estimated frequency of visits appears to build a momentum that carries through the critical period of habit formation.

When comparing these self-reported variables to the direct intervention variables analyzed earlier, it is evident that while the effect of most interventions peaks early and diminishes over time, self-reported variables such as form level, experience level, and estimated visit frequency show a steady increase in their influence. This indicates that intrinsic factors related to the members’ prior fitness levels and behaviors play a crucial role in sustaining long-term engagement with exercise routines. These findings suggest that while interventions can be effective in jump-starting habit formation, the long-term success of these habits is significantly bolstered by the members’ initial fitness levels and commitment to regular exercise.

\section{Discussion}\label{sec4}
This study offers a detailed examination of the factors that influence the formation of consistent exercise habits among gym members. By applying methodologies such as causal inference analysis and cluster-based segmentation, our findings provide nuanced insights that can inform the design of more effective interventions to promote regular physical activity.

One of the key distinctions of our study lies in the development and application of a survival metric to define and measure habit formation. This metric allowed us to capture the persistence of gym attendance over time more precisely than traditional methods. Previous studies primarily relied on gym equipment data \citep{harris19}. In contrast, here we offer a survival metric that identifies critical milestones in the early weeks of gym membership specifically, the 6-week and 17-week marks where the likelihood of maintaining regular attendance significantly declines using comprehensive visitation data. This approach provided a structured framework for understanding when and how gym habits are most vulnerable, thus offering clear targets for intervention.

Furthermore, the causal inference analysis in our study marked a significant advancement in understanding the causal impact of specific gym-related interventions on habit formation. Traditional observational studies often encountered challenges with confounding factors, making it difficult to accurately determine the true effects of various interventions \citep{fischer2008, burke2005}. In contrast, our use of propensity score matching (PSM) allowed us to more effectively isolate the effects of interventions such as group lessons, personal training sessions, visiting different clubs, and invitation credit usage. The analysis revealed that while interventions like personal training consistently demonstrated strong and sustained positive effects on habit formation, other interventions such as group lessons showed diminishing returns over time. However, it is important to highlight that interventions such as visiting different clubs and the use of invitation credits exhibited a more stable impact, with little decline over time. This suggests that while some social and motivational interventions may have temporally limited effectiveness, personalized and varied experiences like those offered through personal training and the opportunity to explore different gym environments maintain their influence over a longer duration. This finding shows the importance of offering a diverse range of interventions to cater to different aspects of habit formation, a detail that has been underexplored in previous studies, particularly those focusing on smaller social exercise settings \citep{wayment}.

Our cluster analysis also introduced a novel way to segment gym members based on their visit patterns, allowing for a more targeted examination of how different subgroups respond to various interventions. Unlike previous studies that often treated gym-goers as a homogeneous group, our segmentation into morning, noon, afternoon, evening, and night clusters revealed significant differences in engagement and responsiveness to interventions. For example, we found that the morning cluster, the peak-time hours for older adults, benefited most from structured and personalized interventions, such as personal training and diverse group lessons. In contrast, the afternoon cluster, dominated by younger adults, responded more positively to interventions that offered variety and novelty, such as visiting different clubs. This highlights the importance of designing interventions that are not only personalized but also adaptable to the specific preferences and schedules of different gym-goer demographics.

Additionally, the analysis of self-reported variables, such as form level, prior experience with fitness, and estimated visit frequency, provided further insights into the intrinsic factors that contribute to habit formation. Unlike the direct interventions, which often showed diminishing returns over time, these self-reported variables exhibited a steadily increasing impact on the likelihood of maintaining consistent exercise habits. This underscores the importance of intrinsic motivation and baseline fitness levels in sustaining long-term engagement.

In synthesizing these findings, it becomes evident that successful habit formation in a gym setting requires a multi-dimensional approach that combines both extrinsic and intrinsic factors. The survival metric highlights the critical periods where interventions can be most effective, while the causal inference analysis provides robust evidence of the causal impact of specific interventions. The cluster analysis underscores the need for tailored interventions that cater to the distinct needs of different member groups, and the role of self-reported variables highlights the value of supporting members' intrinsic motivation.

Building on the insights gained from this study, several avenues for future research and practical implementation can be explored. Firstly, this research aligns with several United Nations Sustainable Development Goals (SDGs) \citep{sdg}, particularly SDG 3 (Good Health and Well-being) and SDG 11 (Sustainable Cities and Communities). Promoting consistent exercise habits contributes to improved public health and well-being, supporting the development of healthier and more resilient communities. Encouraging sustained exercise routines, fitness centers, and public health initiatives can help reduce the incidence of these diseases, contributing to SDG 3’s target of reducing premature mortality from NCDs by one-third by 2030 \citep{sdg}.

Secondly, integrating wearable technology and fitness-tracking apps can provide real-time feedback and support a sense of accountability, which are critical components in sustaining exercise habits. Studies have shown that real-time feedback from wearable devices can significantly boost motivation and adherence to physical activity programs by reinforcing positive behaviors and helping users set realistic goals \citep{patel2015}. By providing continuous monitoring and personalized insights, these digital tools can help individuals stay on track with their fitness routines, adjusting their goals and strategies as needed to maintain engagement over time. This approach could be particularly effective when integrated with existing gym-based interventions, offering a hybrid model that utilizes both the precision of digital health tools and the motivational support of personal trainers.

Moreover, strengthening the social aspects of fitness through community-building activities and group challenges can enhance motivation. Investigating the role of social support networks and peer influence in habit formation can provide further insights into effective strategies, especially given our findings on the temporal limitations of social interventions like group lessons. Additionally, conducting long-term studies to monitor the persistence of exercise habits and the effectiveness of interventions over extended periods will provide deeper insights into the dynamics of habit formation and maintenance. Longitudinal studies can help identify the critical factors that contribute to sustained engagement and inform the design of more effective interventions.

Finally, developing comprehensive intervention models that integrate psychological, social, and environmental factors can offer a more holistic approach to promoting regular exercise. Collaborating with behavioral scientists and public health experts can enhance the effectiveness of these models, ensuring that they address the multi-dimensional nature of habit formation. Additionally, machine learning can identify context variables associated with habit formation, which can inform targeted interventions to increase gym attendance, thereby optimizing the personalization and efficacy of these interventions \citep{buyalskaya}.

In conclusion, this study provides valuable insights into the factors that influence the formation of consistent exercise habits. By integrating causal inference techniques and focusing on both structured interventions and individual characteristics, we offer practical recommendations for gyms and health organizations. Future research should continue to explore innovative strategies and technologies to support sustained physical activity, contributing to improved public health and well-being in alignment with the Sustainable Development Goals.

\section{Data and Code Availability}
All data used in this study were fully anonymized to safeguard personal information in compliance with Türkiye's Personal Data Protection Law (KVKK) \citep{kvkk2024}. This anonymization ensures the privacy of participants while allowing for a comprehensive analysis of gym usage trends and behaviors without compromising individual confidentiality. Due to these constraints we are not able to share any individual level data. However, aggregate statistics are available to reproduce results of the analysis in our GitHub repository: \href{https://github.com/egedemirci/From-Occasional-to-Steady-Habit-Formation-Insights-From-a-Comprehensive-Fitness-Study}{https://github.com/egedemirci/From-Occasional-to-Steady-Habit-Formation-Insights-From-a-Comprehensive-Fitness-Study}.

\section{Acknowledgements}
We would like to express our heartfelt gratitude to Mümtaz Demirci, co-CEO of Mars Athletic Club, for his contributions to the study design and for interpreting the data. We also thank Mehmet Cenk Bursalı and Emre Çakmak for their insights. Furthermore, we extend our thanks to the members of the VRL lab for discussion.

\clearpage

\appendix
\section{}\label{appendix:A}
\renewcommand{\thefigure}{A\arabic{figure}}
\setcounter{figure}{0}
We hypothesized that users may differ in terms of their exercise habits, which we attributed to various variables. To test this idea, we decided to start small and cluster one of the clubs. The idea to start at a particular club was provided to us by our correspondents at Mars Athletic Club. They mentioned that one of the clubs resembled the overall population closely in terms of age and gender distribution and had a high number of people, which would give us enough data to analyze and feed into the NMF algorithm to produce meaningful results. Below in Fig\ref{fig:age-and-gender-dist}, the age and gender distributions of this club are showed. Since these statistics showed little deviation from overall, we went ahead and applied the same procedures to all customers. This first experiment was essential in the process of clustering, as it provided us with what to expect from other clubs as well and shaped the later iterations to come up with the final results.

\begin{figure}[H]
\centering
\includegraphics[width=\textwidth]{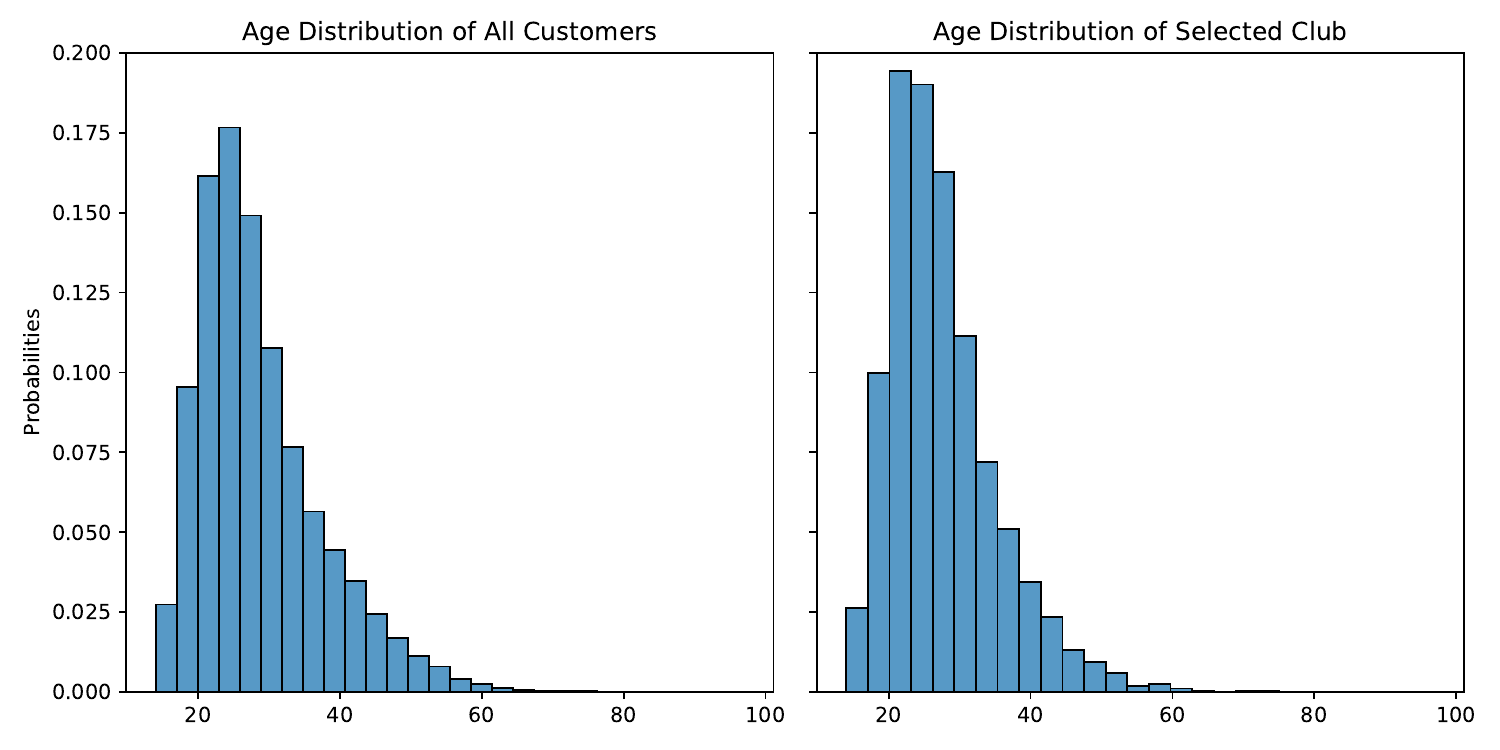}
\includegraphics[width=\linewidth]{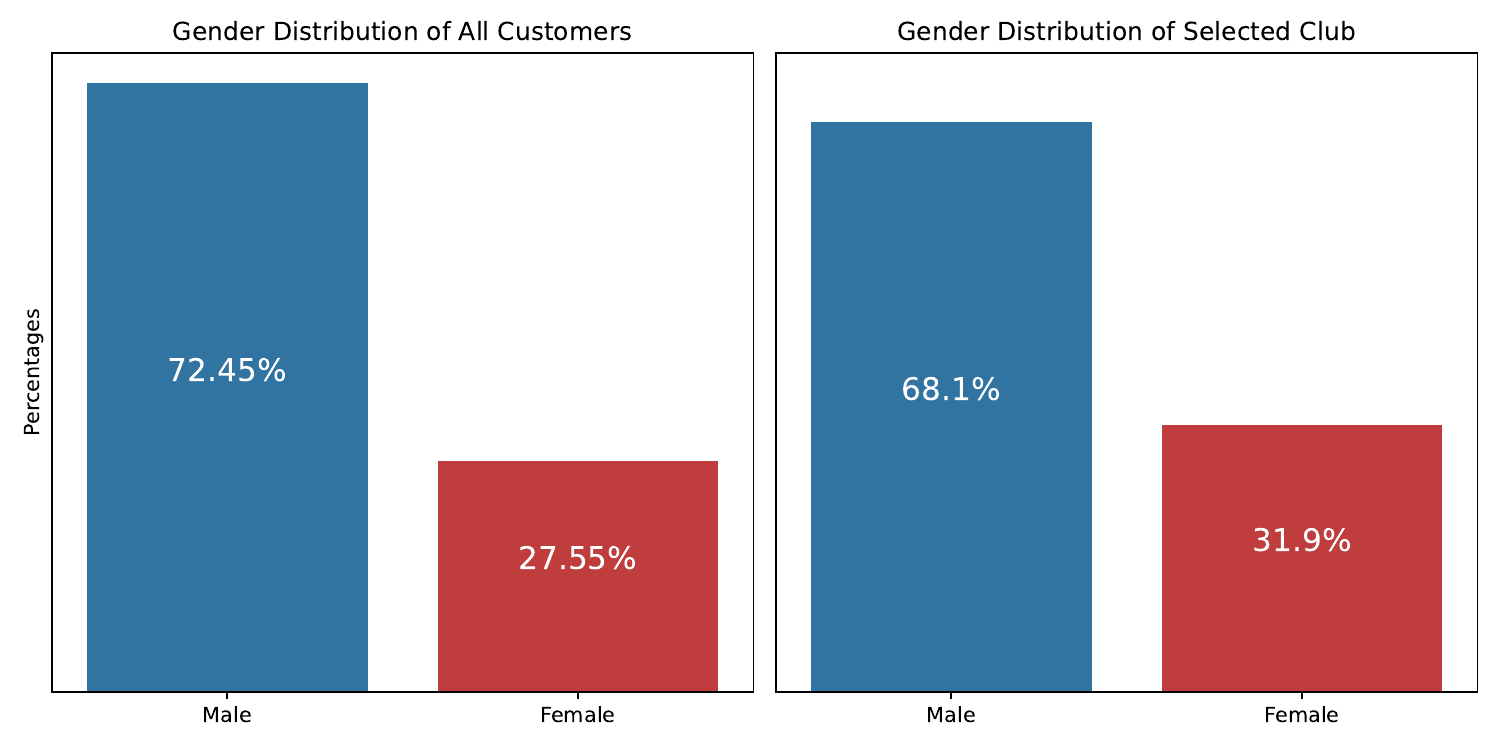}
\caption{Age and gender distributions.}
\label{fig:age-and-gender-dist}
\end{figure}

\newpage
\section{}\label{appendix:B}
To cluster all customers, we needed to vectorize their visits so that we could feed them into the NMF algorithm. Here, we show how the vectorization of all customers' visits were achieved. All customers have their own unique dictionary which represents their weekly visit habits. The dictionaries are then fed into the NMF to produce the distances to cluster centers. Since the gyms worked less hours in the weekends, we set the thresholds differently.

\begin{algorithm}[H]
\caption{Pseudocode for vectorization script.}
\label{alg:vec-script}
\begin{algorithmic}[1]
\State \textbf{Initialize} \texttt{hours} as a dictionary with keys as hourly intervals from 6:00 to 24:00 and values set to 0.
\State \textbf{Initialize} \texttt{hours\_template} as a dictionary with keys as days of the week (0-6) and values as copies of \texttt{hour\_temp}.

\Procedure{PreparePatterns}{\texttt{user\_data, entry, exit, day}}
    \If{\texttt{day} is a weekend day}
        \For{\texttt{hour} in range from \texttt{entry} to min(\texttt{exit}+1, 20)}
            \State Increment \texttt{user\_data[day][f"{hour}:00-{hour+1}:00"]} by 1
        \EndFor
    \Else
        \For{\texttt{hour} in range from \texttt{entry} to min(\texttt{exit}+1, 23)}
            \State Increment \texttt{user\_data[day][f"{hour}:00-{hour+1}:00"]} by 1
        \EndFor
    \EndIf
\EndProcedure
\end{algorithmic}
\end{algorithm}

\section{}\label{appendix:C}
\renewcommand{\thefigure}{C\arabic{figure}}
\setcounter{figure}{0}

When considering the conditions under which a survival streak is preserved or lost, it is essential to recognize the concept of tolerance for occasional absences. Specifically, the metric permits members to miss gym visits for up to one week without breaking their survival streak. This allowance recognizes that individuals may encounter situations that temporarily restrict their ability to attend the gym, such as travel obligations, illness, or exceptionally demanding work or personal commitments. As a result of this tolerance period, if a member fails to visit the gym for two consecutive weeks, their survival streak is broken. This threshold signifies a departure from consistent gym attendance habits, indicating a potential disruption in forming long-term exercise routines.

\begin{figure}[H]
\centering
\includegraphics[width=\textwidth]{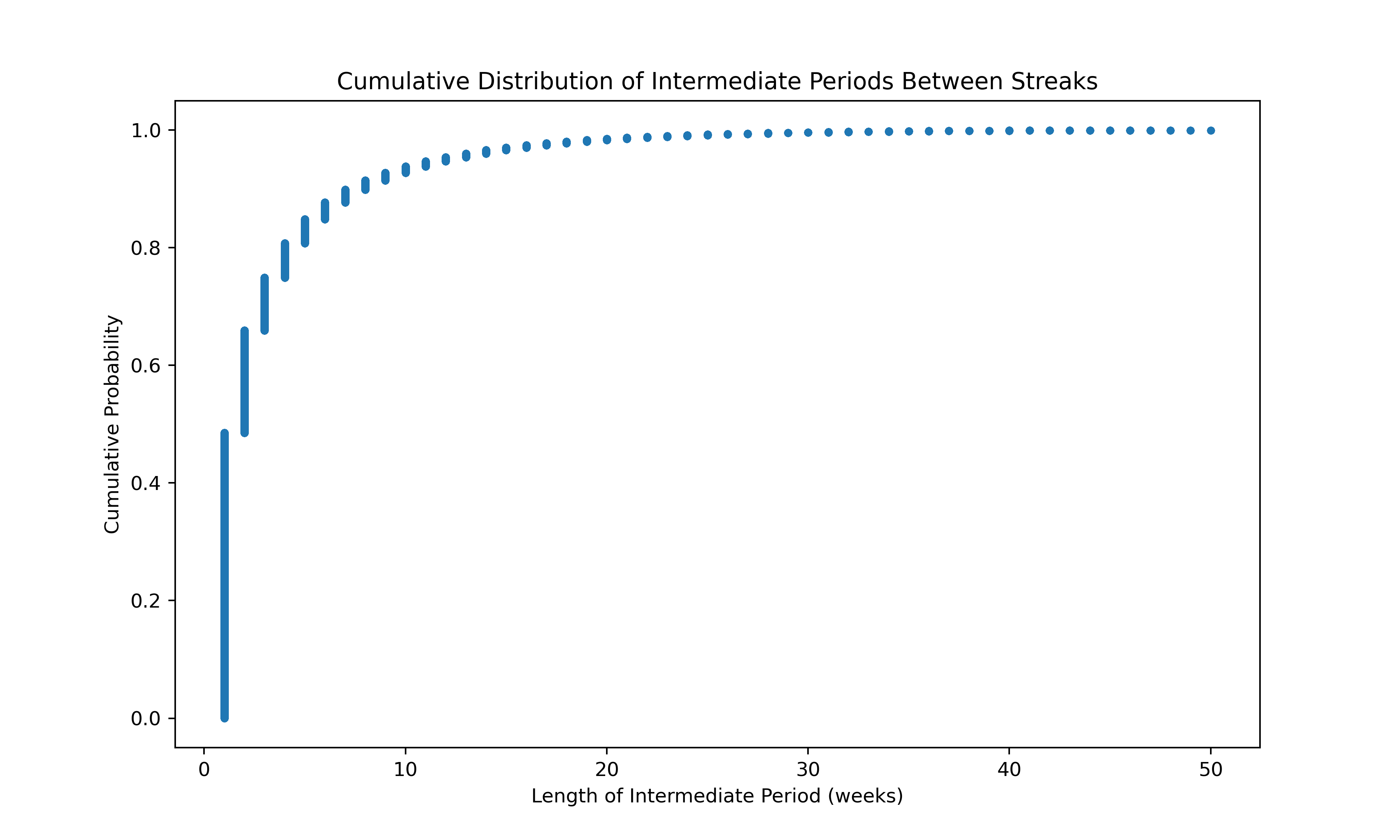}
\caption{Cumulative distribution of the intermediate periods between streaks.}
\label{fig:intermediate_gaps}
\end{figure}

To determine that a two-week absence is the threshold for breaking the survival streak, we analyzed all intermediate periods between consecutive gym attendance periods, as shown in Figure \ref{fig:intermediate_gaps}. The Figure displays the cumulative distribution of each intermediate gap between streaks, with 50\% of all intermediate gaps being one week long. Therefore, we decided to tolerate these one-week gaps between streaks.

\section{}\label{appendix:D}
\renewcommand{\thefigure}{D\arabic{figure}}
\setcounter{figure}{0}

\begin{figure}[H]
\centering
\includegraphics[width=\textwidth]{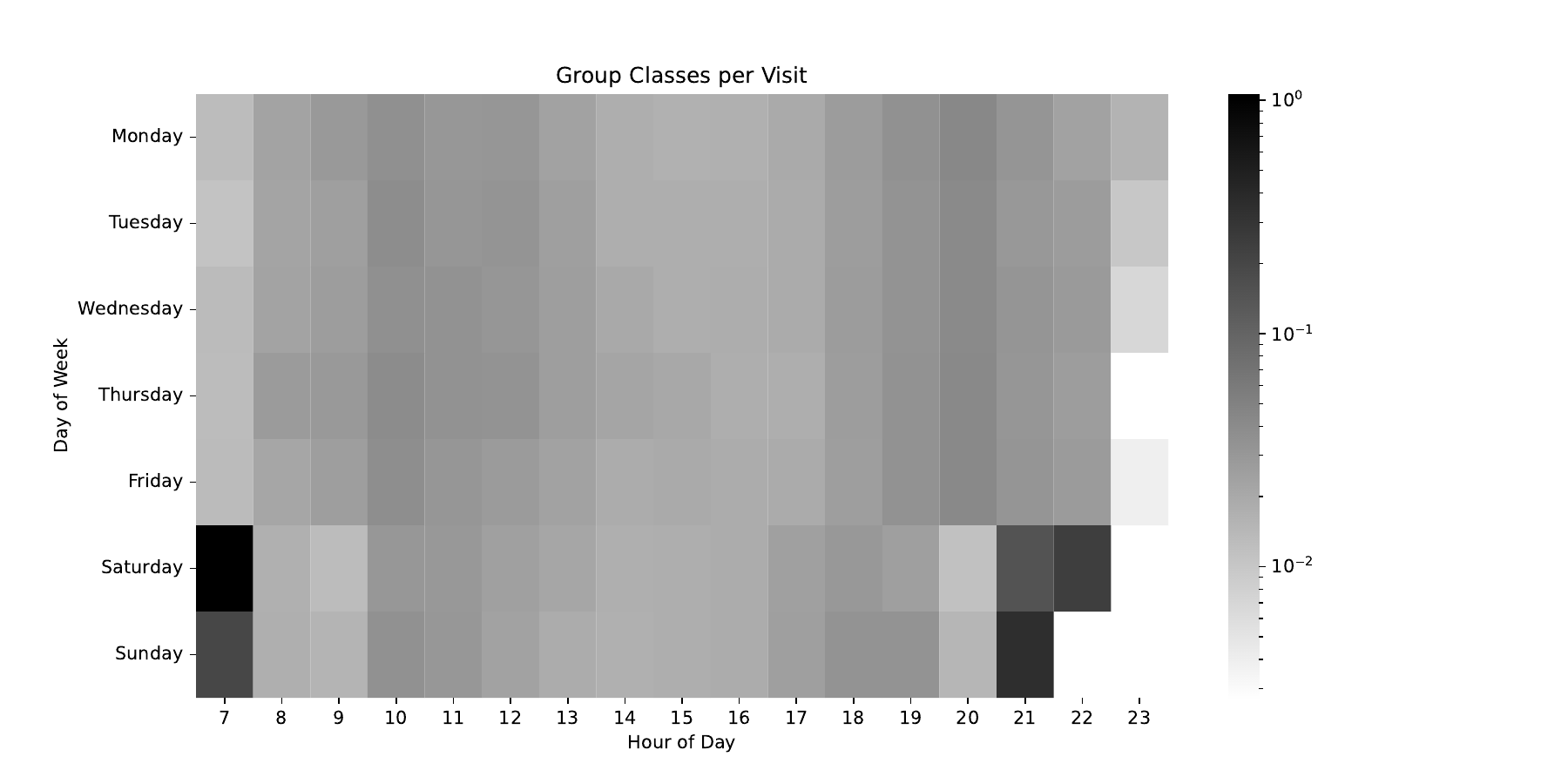}
\caption{Group classes opened per visit.}
\label{fig:group_class_per_vst}
\end{figure}

\section{}\label{appendix:E}
\renewcommand{\thefigure}{E\arabic{figure}}
\setcounter{figure}{0}

\begin{figure}[H]
    \centering
    \begin{subfigure}[b]{0.3925\textwidth}
        \centering
        \includegraphics[width=\textwidth]{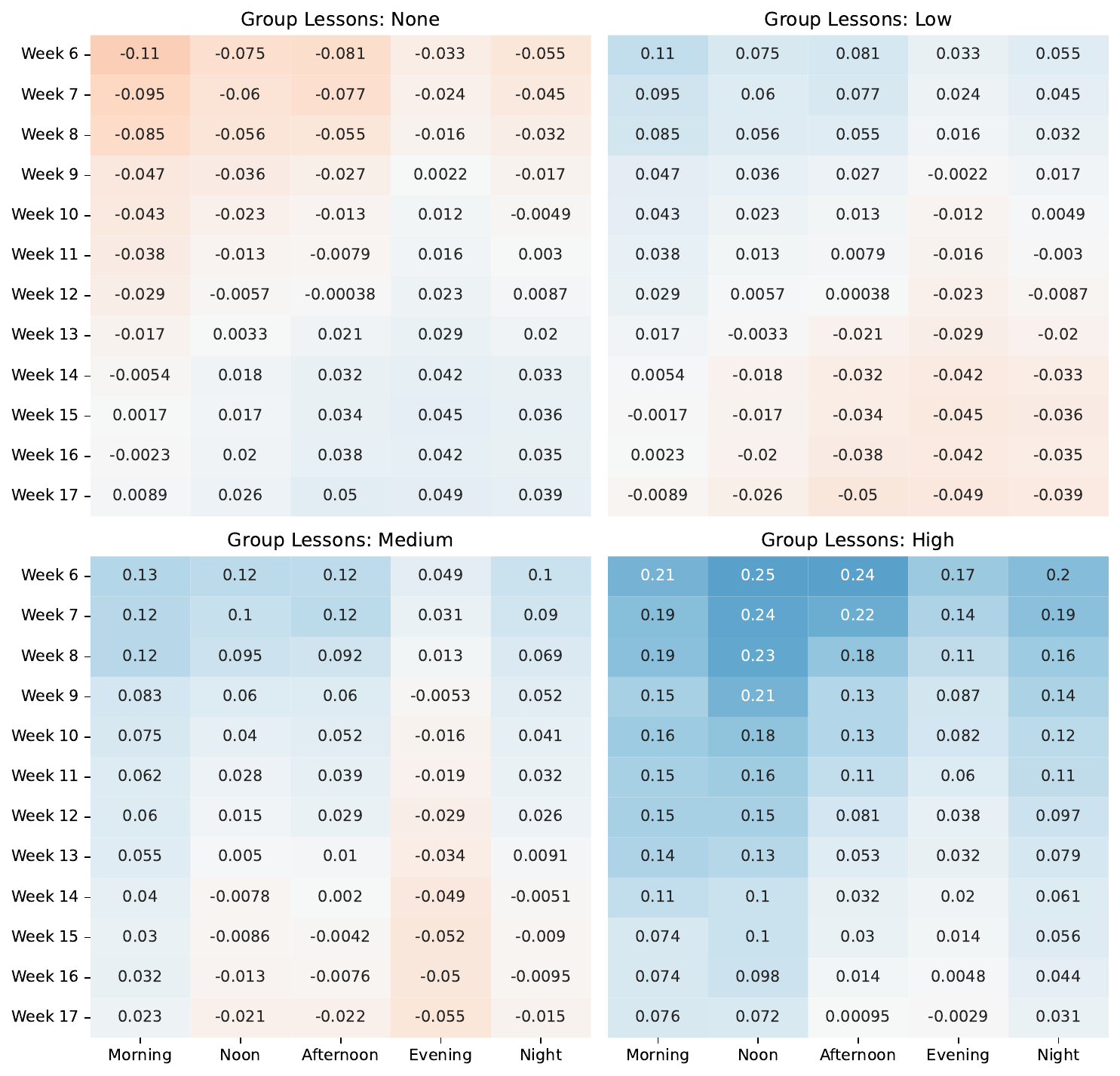}
        \caption{Group Lessons}
        \label{fig:ind1}
    \end{subfigure}
    \hfill
    \begin{subfigure}[b]{0.3925\textwidth}
        \centering
        \includegraphics[width=\textwidth]{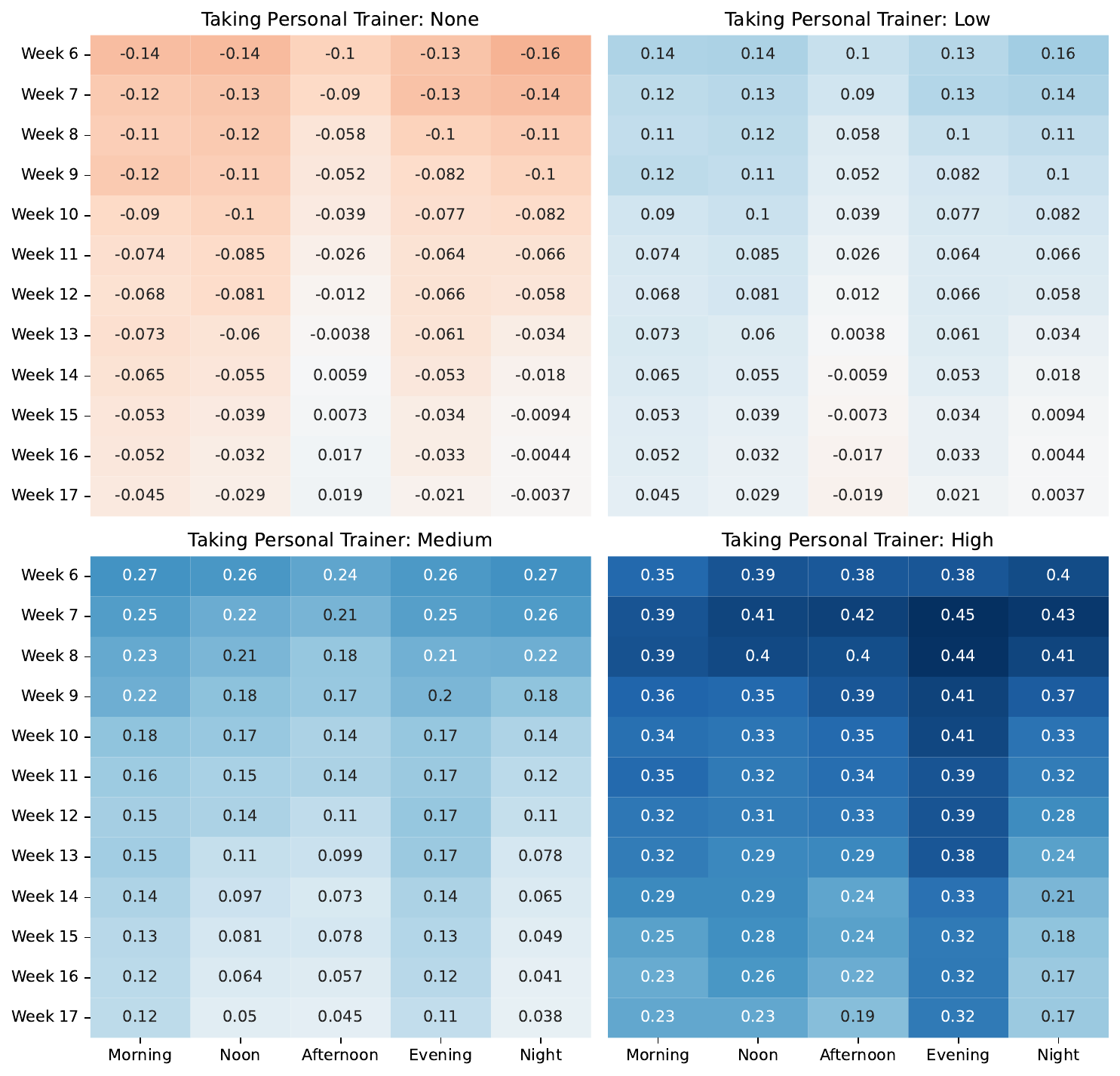}
        \caption{Taking Personal Trainer}
        \label{fig:ind2}
    \end{subfigure}

    \begin{subfigure}[b]{0.3925\textwidth}
        \centering
        \includegraphics[width=\textwidth]{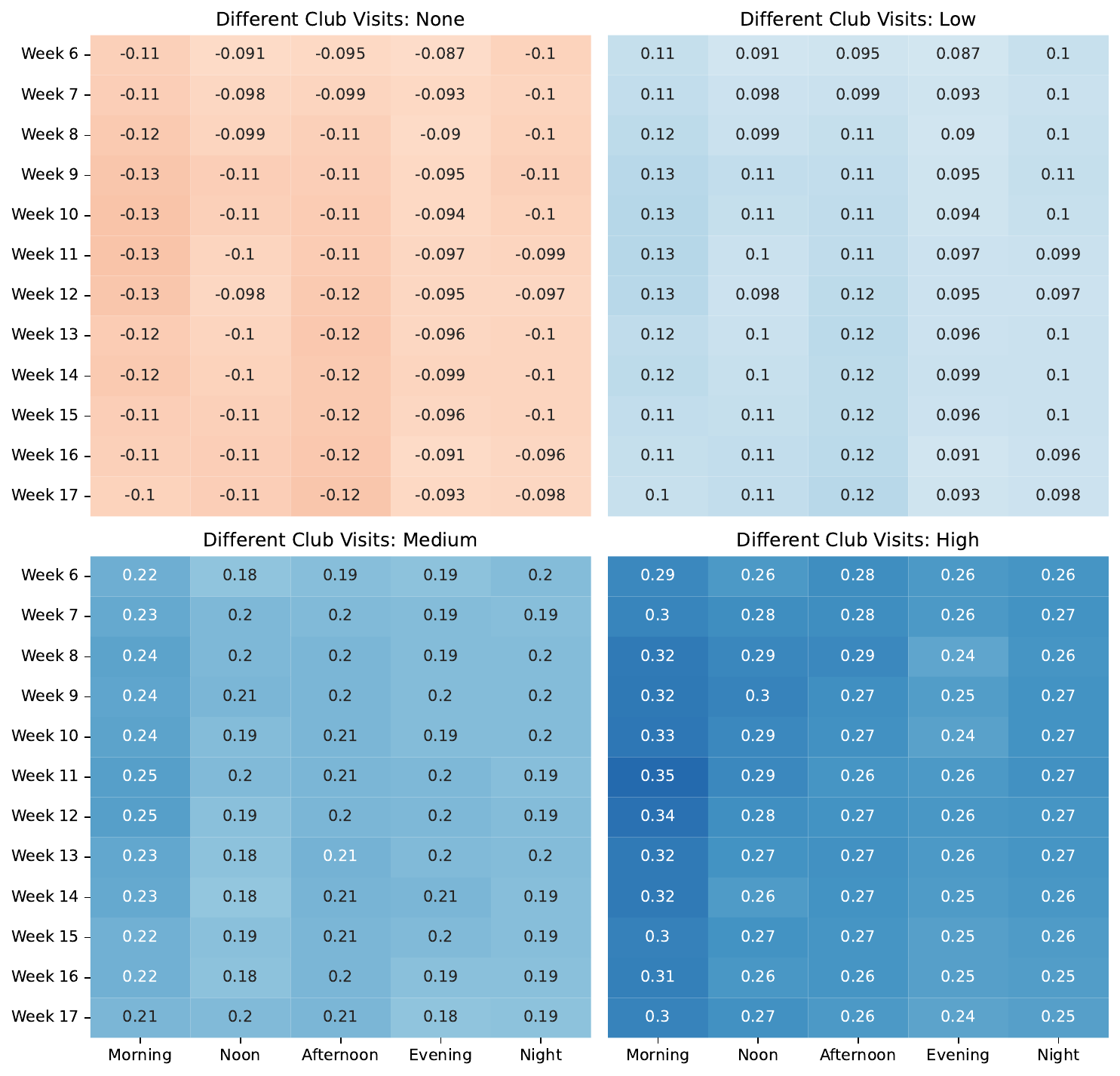}
        \caption{Different Club Visits}
        \label{fig:ind3}
    \end{subfigure}
    \hfill
    \begin{subfigure}[b]{0.3925\textwidth}
        \centering
        \includegraphics[width=\textwidth]{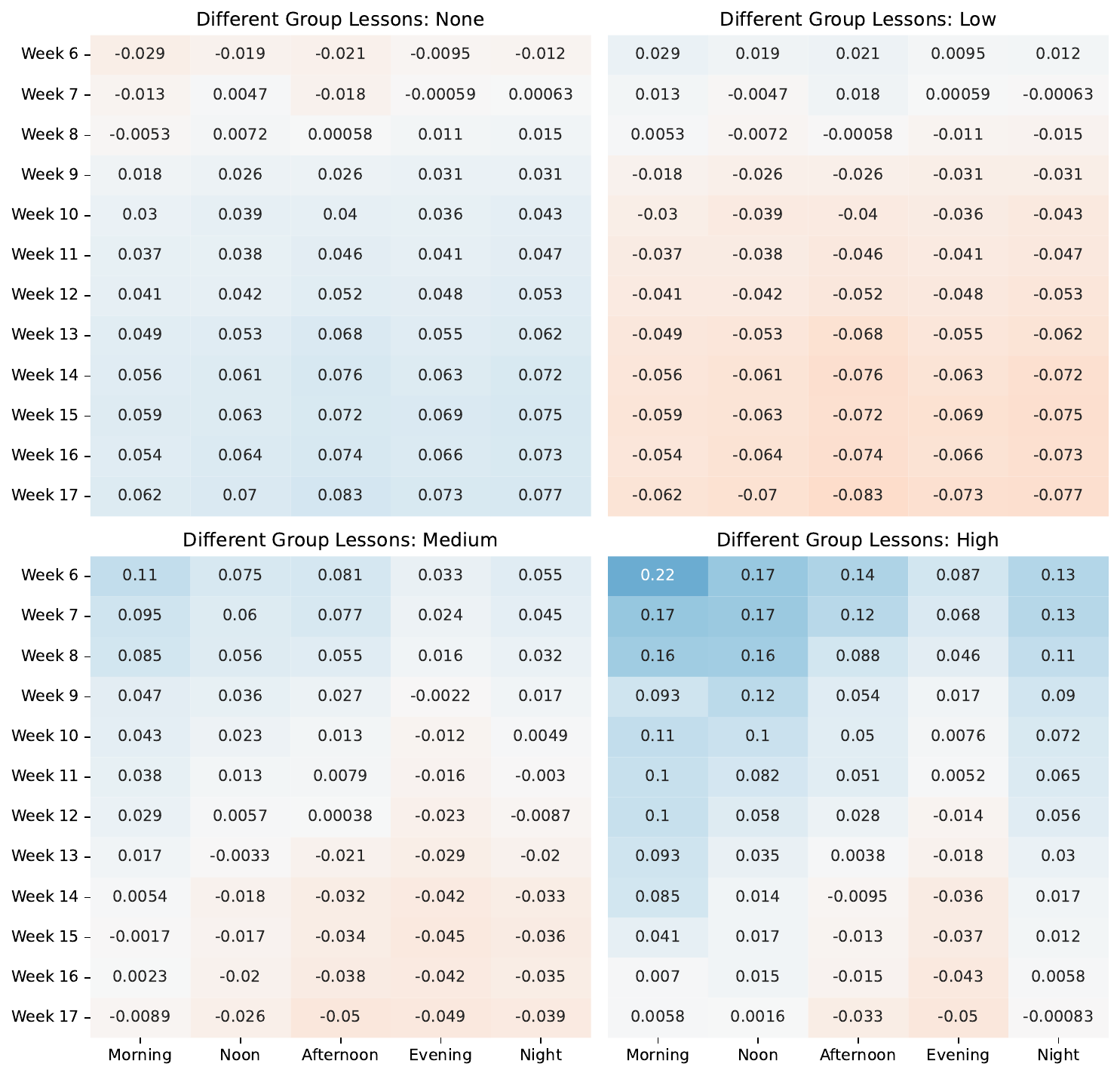}
        \caption{Different Group Lessons}
        \label{fig:ind4}
    \end{subfigure}

    \begin{subfigure}[b]{0.3925\textwidth}
        \centering
        \includegraphics[width=\textwidth]{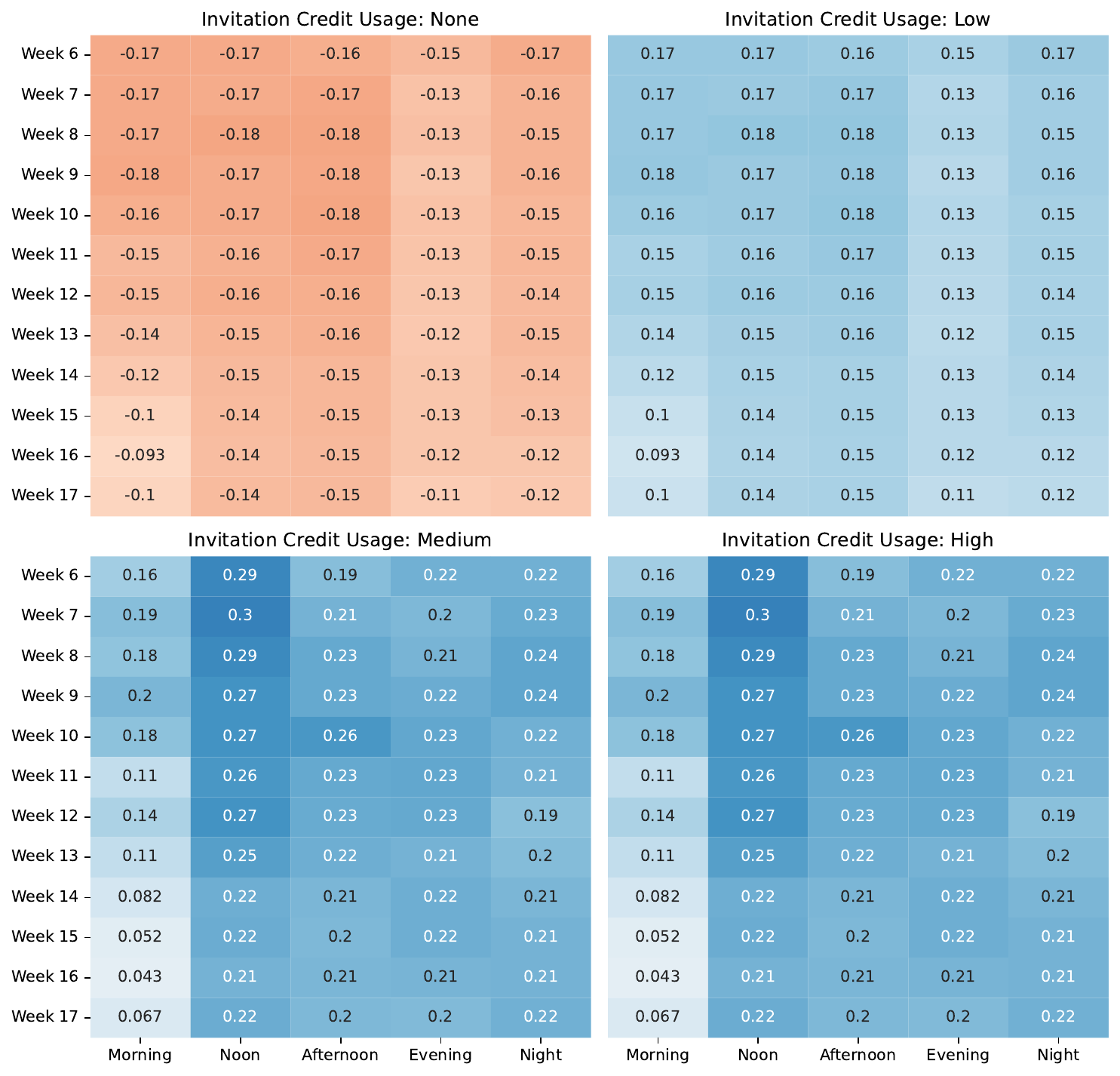}
        \caption{Invitation Credit Usage}
        \label{fig:ind5}
    \end{subfigure}
    \caption{Estimated effects of interventions on habit formation (individual clusters).}
    \label{fig:intervention-effects}
\end{figure}

%%\bibliographystyle{elsarticle-num}
%\bibliographystyle{elsarticle-harv}
%\bibliography{sn-bibliography}

\begin{thebibliography}{30}
\expandafter\ifx\csname natexlab\endcsname\relax\def\natexlab#1{#1}\fi
\providecommand{\url}[1]{\texttt{#1}}
\providecommand{\href}[2]{#2}
\providecommand{\path}[1]{#1}
\providecommand{\DOIprefix}{doi:}
\providecommand{\ArXivprefix}{arXiv:}
\providecommand{\URLprefix}{URL: }
\providecommand{\Pubmedprefix}{pmid:}
\providecommand{\doi}[1]{\href{http://dx.doi.org/#1}{\path{#1}}}
\providecommand{\Pubmed}[1]{\href{pmid:#1}{\path{#1}}}
\providecommand{\bibinfo}[2]{#2}
\ifx\xfnm\relax \def\xfnm[#1]{\unskip,\space#1}\fi
%Type = Article
\bibitem[{Barkley et~al.(2020)Barkley, Lepp, Santo, Glickman and Dowdell}]{barkley}
\bibinfo{author}{Barkley, J.E.}, \bibinfo{author}{Lepp, A.}, \bibinfo{author}{Santo, A.}, \bibinfo{author}{Glickman, E.}, \bibinfo{author}{Dowdell, B.}, \bibinfo{year}{2020}.
\newblock \bibinfo{title}{The relationship between fitness app use and physical activity behavior is mediated by exercise identity}.
\newblock \bibinfo{journal}{Computers in Human Behavior} \bibinfo{volume}{108}, \bibinfo{pages}{106313}.
\newblock \URLprefix \url{https://www.sciencedirect.com/science/article/pii/S0747563220300674}, \DOIprefix\doi{https://doi.org/10.1016/j.chb.2020.106313}.
%Type = Article
\bibitem[{Burke et~al.(2005)Burke, Carron, Eys, Ntoumanis and Estabrooks}]{burke2005}
\bibinfo{author}{Burke, S.}, \bibinfo{author}{Carron, A.}, \bibinfo{author}{Eys, M.}, \bibinfo{author}{Ntoumanis, N.}, \bibinfo{author}{Estabrooks, P.}, \bibinfo{year}{2005}.
\newblock \bibinfo{title}{Group versus individual approach? a meta-analysis of the effectiveness of interventions to promote physical activity}.
\newblock \bibinfo{journal}{International Review of Sport and Exercise Psychology} \bibinfo{volume}{2}, \bibinfo{pages}{13}.
\newblock \DOIprefix\doi{10.53841/bpssepr.2006.2.1.13}.
%Type = Article
\bibitem[{Buyalskaya et~al.(2023)Buyalskaya, Ho, Milkman, Li, Duckworth and Camerer}]{buyalskaya}
\bibinfo{author}{Buyalskaya, A.}, \bibinfo{author}{Ho, H.}, \bibinfo{author}{Milkman, K.L.}, \bibinfo{author}{Li, X.}, \bibinfo{author}{Duckworth, A.L.}, \bibinfo{author}{Camerer, C.}, \bibinfo{year}{2023}.
\newblock \bibinfo{title}{What can machine learning teach us about habit formation? evidence from exercise and hygiene}.
\newblock \bibinfo{journal}{Proc. Natl. Acad. Sci. U. S. A.} \bibinfo{volume}{120}, \bibinfo{pages}{e2216115120}.
%Type = Inbook
\bibitem[{Ding et~al.(2005)Ding, He and Simon}]{ding2005}
\bibinfo{author}{Ding, C.}, \bibinfo{author}{He, X.}, \bibinfo{author}{Simon, H.D.}, \bibinfo{year}{2005}.
\newblock \bibinfo{title}{On the Equivalence of Nonnegative Matrix Factorization and Spectral Clustering}. \bibinfo{publisher}{Society for Industrial and Applied Mathematics}.
\newblock pp. \bibinfo{pages}{606--610}.
\newblock \URLprefix \url{https://epubs.siam.org/doi/abs/10.1137/1.9781611972757.70}, \DOIprefix\doi{10.1137/1.9781611972757.70}, \href{http://arxiv.org/abs/https://epubs.siam.org/doi/pdf/10.1137/1.9781611972757.70}{{\tt arXiv:https://epubs.siam.org/doi/pdf/10.1137/1.9781611972757.70}}.
%Type = Article
\bibitem[{Fischer and Bryant(2008)}]{fischer2008}
\bibinfo{author}{Fischer, D.}, \bibinfo{author}{Bryant, J.}, \bibinfo{year}{2008}.
\newblock \bibinfo{title}{Effect of certified personal trainer services on stage of exercise behavior and exercise mediators in female college students}.
\newblock \bibinfo{journal}{Journal of American college health : J of ACH} \bibinfo{volume}{56}, \bibinfo{pages}{369--76}.
\newblock \DOIprefix\doi{10.3200/JACH.56. 44.369-376}.
%Type = Article
\bibitem[{Galvim et~al.(2019)Galvim, Oliveira, Martins, Vieira, Cerri, Cezar, Pedroso and Angelica~de Oliveira~Gomes}]{galvim}
\bibinfo{author}{Galvim, A.}, \bibinfo{author}{Oliveira, I.}, \bibinfo{author}{Martins, T.}, \bibinfo{author}{Vieira, L.}, \bibinfo{author}{Cerri, N.}, \bibinfo{author}{Cezar, N.}, \bibinfo{author}{Pedroso, R.}, \bibinfo{author}{Angelica~de Oliveira~Gomes, G.}, \bibinfo{year}{2019}.
\newblock \bibinfo{title}{Adherence, adhesion, and dropout reasons of a physical activity program in a high social vulnerability context}.
\newblock \bibinfo{journal}{Journal of Physical Activity and Health} \bibinfo{volume}{16}, \bibinfo{pages}{1--8}.
\newblock \DOIprefix\doi{10.1123/jpah.2017-0606}.
%Type = Article
\bibitem[{Gardner et~al.(2012)Gardner, Lally and Wardle}]{gardner2}
\bibinfo{author}{Gardner, B.}, \bibinfo{author}{Lally, P.}, \bibinfo{author}{Wardle, J.}, \bibinfo{year}{2012}.
\newblock \bibinfo{title}{Making health habitual: The psychology of 'habit-formation' and general practice}.
\newblock \bibinfo{journal}{The British journal of general practice : the journal of the Royal College of General Practitioners} \bibinfo{volume}{62}, \bibinfo{pages}{664--6}.
\newblock \DOIprefix\doi{10.3399/bjgp12X659466}.
%Type = Article
\bibitem[{Gardner et~al.(2022)Gardner, Rebar and and}]{gardner}
\bibinfo{author}{Gardner, B.}, \bibinfo{author}{Rebar, A.L.}, \bibinfo{author}{and, P.L.}, \bibinfo{year}{2022}.
\newblock \bibinfo{title}{How does habit form? guidelines for tracking real-world habit formation}.
\newblock \bibinfo{journal}{Cogent Psychology} \bibinfo{volume}{9}, \bibinfo{pages}{2041277}.
\newblock \URLprefix \url{https://doi.org/10.1080/23311908.2022.2041277}, \DOIprefix\doi{10.1080/23311908.2022.2041277}, \href{http://arxiv.org/abs/https://doi.org/10.1080/23311908.2022.2041277}{{\tt arXiv:https://doi.org/10.1080/23311908.2022.2041277}}.
%Type = Article
\bibitem[{Harris and Kessler(2019)}]{harris19}
\bibinfo{author}{Harris, M.C.}, \bibinfo{author}{Kessler, L.M.}, \bibinfo{year}{2019}.
\newblock \bibinfo{title}{Habit formation and activity persistence: Evidence from gym equipment}.
\newblock \bibinfo{journal}{Journal of Economic Behavior \& Organization} \bibinfo{volume}{166}, \bibinfo{pages}{688--708}.
\newblock \URLprefix \url{https://www.sciencedirect.com/science/article/pii/S0167268119302586}, \DOIprefix\doi{https://doi.org/10.1016/j.jebo.2019.08.010}.
%Type = Article
\bibitem[{Huang et~al.(2022)Huang, Sun and Jiang}]{huangg}
\bibinfo{author}{Huang, G.}, \bibinfo{author}{Sun, M.}, \bibinfo{author}{Jiang, L.C.}, \bibinfo{year}{2022}.
\newblock \bibinfo{title}{Core social network size is associated with physical activity participation for fitness app users: The role of social comparison and social support}.
\newblock \bibinfo{journal}{Computers in Human Behavior} \bibinfo{volume}{129}.
\newblock \URLprefix \url{https://doi.org/10.1016/j.chb.2021.107169}, \DOIprefix\doi{10.1016/j.chb.2021.107169}.
%Type = Article
\bibitem[{van Kasteren et~al.(2020)van Kasteren, Lewis and Maeder}]{kasteren}
\bibinfo{author}{van Kasteren, Y.}, \bibinfo{author}{Lewis, L.}, \bibinfo{author}{Maeder, A.}, \bibinfo{year}{2020}.
\newblock \bibinfo{title}{Office-based physical activity: mapping a social ecological model approach against com-b}.
\newblock \bibinfo{journal}{BMC Public Health} \bibinfo{volume}{20}.
\newblock \DOIprefix\doi{10.1186/s12889-020-8280-1}.
%Type = Article
\bibitem[{Kaushal and Rhodes(2015)}]{kaushal2015}
\bibinfo{author}{Kaushal, N.}, \bibinfo{author}{Rhodes, R.E.}, \bibinfo{year}{2015}.
\newblock \bibinfo{title}{Exercise habit formation in new gym members: A longitudinal study}.
\newblock \bibinfo{journal}{Journal of Behavioral Medicine} \bibinfo{volume}{38}, \bibinfo{pages}{652--663}.
\newblock \DOIprefix\doi{10.1007/s10865-015-9640-7}.
%Type = Techreport
\bibitem[{Kim and Park(2008)}]{Kim2008SparseNM}
\bibinfo{author}{Kim, J.}, \bibinfo{author}{Park, H.}, \bibinfo{year}{2008}.
\newblock \bibinfo{title}{Sparse nonnegative matrix factorization for clustering}.
\newblock \bibinfo{type}{Technical Report}. Georgia Institute of Technology.
%Type = Article
\bibitem[{Lally et~al.(2010)Lally, Jaarsveld, Potts and Wardle}]{lally}
\bibinfo{author}{Lally, P.}, \bibinfo{author}{Jaarsveld, C.V.}, \bibinfo{author}{Potts, H.}, \bibinfo{author}{Wardle, J.}, \bibinfo{year}{2010}.
\newblock \bibinfo{title}{How are habits formed: Modelling habit formation in the real world}.
\newblock \bibinfo{journal}{European Journal of Social Psychology} \bibinfo{volume}{40}, \bibinfo{pages}{998--1009}.
%Type = Article
\bibitem[{Lee et~al.(2012)Lee, Shiroma, Lobelo, Puska, Blair, Katzmarzyk and Group}]{lee2012effect}
\bibinfo{author}{Lee, I.M.}, \bibinfo{author}{Shiroma, E.J.}, \bibinfo{author}{Lobelo, F.}, \bibinfo{author}{Puska, P.}, \bibinfo{author}{Blair, S.N.}, \bibinfo{author}{Katzmarzyk, P.T.}, \bibinfo{author}{Group, L.P.A.S.W.}, \bibinfo{year}{2012}.
\newblock \bibinfo{title}{Effect of physical inactivity on major non-communicable diseases worldwide: an analysis of burden of disease and life expectancy}.
\newblock \bibinfo{journal}{The Lancet} \bibinfo{volume}{380}, \bibinfo{pages}{219--229}.
\newblock \URLprefix \url{https://doi.org/10.1016/S0140-6736(12)61031-9}, \DOIprefix\doi{10.1016/S0140-6736(12)61031-9}.
%Type = Article
\bibitem[{Li and Meng(2023)}]{limeng}
\bibinfo{author}{Li, L.}, \bibinfo{author}{Meng, J.}, \bibinfo{year}{2023}.
\newblock \bibinfo{title}{Network effects on physical activity through interpersonal vs. masspersonal communication with the core and acquaintance networks}.
\newblock \bibinfo{journal}{Computers in Human Behavior} \bibinfo{volume}{141}, \bibinfo{pages}{107594}.
\newblock \URLprefix \url{https://www.sciencedirect.com/science/article/pii/S0747563222004149}, \DOIprefix\doi{https://doi.org/10.1016/j.chb.2022.107594}.
%Type = Article
\bibitem[{Michie et~al.(2011)Michie, van Stralen and West}]{michie}
\bibinfo{author}{Michie, S.}, \bibinfo{author}{van Stralen, M.}, \bibinfo{author}{West, R.}, \bibinfo{year}{2011}.
\newblock \bibinfo{title}{The behaviour change wheel: a new method for characterising and designing behaviour change interventions}.
\newblock \bibinfo{journal}{Implementation science : IS} \bibinfo{volume}{6}, \bibinfo{pages}{42}.
\newblock \DOIprefix\doi{10.1186/1748-5908-6-42}.
%Type = Misc
\bibitem[{Nations(2015)}]{sdg}
\bibinfo{author}{Nations, U.}, \bibinfo{year}{2015}.
\newblock \bibinfo{title}{Sustainable development goals}.
\newblock \bibinfo{howpublished}{\url{https://sdgs.un.org/goals}}.
%Type = Article
\bibitem[{Nuss and Li(2021)}]{nuss}
\bibinfo{author}{Nuss, K.}, \bibinfo{author}{Li, K.}, \bibinfo{year}{2021}.
\newblock \bibinfo{title}{Motivation for physical activity and physcial activity engagement in current and former wearable fitness tracker users: A mixed-methods examination}.
\newblock \bibinfo{journal}{Computers in Human Behavior} \bibinfo{volume}{121}, \bibinfo{pages}{106798}.
\newblock \URLprefix \url{https://www.sciencedirect.com/science/article/pii/S0747563221001217}, \DOIprefix\doi{https://doi.org/10.1016/j.chb.2021.106798}.
%Type = Article
\bibitem[{Oc and Plangger(2022)}]{gist}
\bibinfo{author}{Oc, Y.}, \bibinfo{author}{Plangger, K.}, \bibinfo{year}{2022}.
\newblock \bibinfo{title}{Gist do it! how motivational mechanisms help wearable users develop healthy habits}.
\newblock \bibinfo{journal}{Computers in Human Behavior} \bibinfo{volume}{128}.
\newblock \URLprefix \url{https://doi.org/10.1016/j.chb.2021.107089}, \DOIprefix\doi{10.1016/j.chb.2021.107089}.
%Type = Article
\bibitem[{Patel et~al.(2015)Patel, Asch and Volpp}]{patel2015}
\bibinfo{author}{Patel, M.}, \bibinfo{author}{Asch, D.}, \bibinfo{author}{Volpp, K.}, \bibinfo{year}{2015}.
\newblock \bibinfo{title}{Wearable devices as facilitators, not drivers, of health behavior change}.
\newblock \bibinfo{journal}{JAMA} \bibinfo{volume}{313}.
\newblock \DOIprefix\doi{10.1001/jama.2014.14781}.
%Type = Misc
\bibitem[{{Personal Data Protection Authority}(2016)}]{kvkk2024}
\bibinfo{author}{{Personal Data Protection Authority}}, \bibinfo{year}{2016}.
\newblock \bibinfo{title}{Personal data protection law}.
\newblock \URLprefix \url{https://www.kvkk.gov.tr/Icerik/6649/Personal-Data-Protection-Law}. \bibinfo{note}{accessed: 2024-08-23}.
%Type = Article
\bibitem[{Sallis et~al.(2006)Sallis, Cervero, Ascher and Henderson}]{sallis}
\bibinfo{author}{Sallis, J.}, \bibinfo{author}{Cervero, R.}, \bibinfo{author}{Ascher, W.}, \bibinfo{author}{Henderson, K.}, \bibinfo{year}{2006}.
\newblock \bibinfo{title}{An ecological approach to creating active living communities}.
\newblock \bibinfo{journal}{ANNUAL REVIEW OF PUBLIC HEALTH} \bibinfo{volume}{27}.
%Type = Article
\bibitem[{Sallis(2018)}]{sallis2018}
\bibinfo{author}{Sallis, J.F.}, \bibinfo{year}{2018}.
\newblock \bibinfo{title}{Needs and challenges related to multilevel interventions: Physical activity examples}.
\newblock \bibinfo{journal}{Health Education \& Behavior} \bibinfo{volume}{45}, \bibinfo{pages}{661--667}.
\newblock \URLprefix \url{https://doi.org/10.1177/1090198118796458}, \DOIprefix\doi{10.1177/1090198118796458}, \href{http://arxiv.org/abs/https://doi.org/10.1177/1090198118796458}{{\tt arXiv:https://doi.org/10.1177/1090198118796458}}. \bibinfo{note}{pMID: 30122086}.
%Type = Misc
\bibitem[{Sharma and Kiciman(2020)}]{dowhy}
\bibinfo{author}{Sharma, A.}, \bibinfo{author}{Kiciman, E.}, \bibinfo{year}{2020}.
\newblock \bibinfo{title}{Dowhy: An end-to-end library for causal inference}.
\newblock \bibinfo{howpublished}{\url{https://arxiv.org/abs/2011.04216}}.
%Type = Article
\bibitem[{Singh et~al.(2024)Singh, Murphy, Maher and Smith}]{singh}
\bibinfo{author}{Singh, B.}, \bibinfo{author}{Murphy, A.}, \bibinfo{author}{Maher, C.}, \bibinfo{author}{Smith, A.}, \bibinfo{year}{2024}.
\newblock \bibinfo{title}{Time to form a habit: A systematic review and meta-analysis of health behaviour habit formation and its determinants}.
\newblock \bibinfo{journal}{Healthcare} \bibinfo{volume}{12}, \bibinfo{pages}{2488}.
\newblock \DOIprefix\doi{10.3390/healthcare12232488}.
%Type = Article
\bibitem[{Wayment and McDonald(2017)}]{wayment}
\bibinfo{author}{Wayment, H.A.}, \bibinfo{author}{McDonald, R.L.}, \bibinfo{year}{2017}.
\newblock \bibinfo{title}{Sharing a personal trainer: Personal and social benefits of individualized, small-group training}.
\newblock \bibinfo{journal}{The Journal of Strength {\&} Conditioning Research} \bibinfo{volume}{31}.
\newblock \URLprefix \url{https://journals.lww.com/nsca-jscr/fulltext/2017/11000/sharing_a_personal_trainer__personal_and_social.24.aspx}.
%Type = Article
\bibitem[{Weyland et~al.(2020)Weyland, Finne, Krell-Roesch and Jekauc}]{weyland}
\bibinfo{author}{Weyland, S.}, \bibinfo{author}{Finne, E.}, \bibinfo{author}{Krell-Roesch, J.}, \bibinfo{author}{Jekauc, D.}, \bibinfo{year}{2020}.
\newblock \bibinfo{title}{(how) does affect influence the formation of habits in exercise?}
\newblock \bibinfo{journal}{Frontiers in Psychology} \bibinfo{volume}{Volume 11 - 2020}.
\newblock \URLprefix \url{https://www.frontiersin.org/journals/psychology/articles/10.3389/fpsyg.2020.578108}, \DOIprefix\doi{10.3389/fpsyg.2020.578108}.
%Type = Article
\bibitem[{Wood and R\"{u}nger(2016)}]{Wood2016}
\bibinfo{author}{Wood, W.}, \bibinfo{author}{R\"{u}nger, D.}, \bibinfo{year}{2016}.
\newblock \bibinfo{title}{Psychology of habit}.
\newblock \bibinfo{journal}{Annual Review of Psychology} \bibinfo{volume}{67}, \bibinfo{pages}{289--314}.
%Type = Misc
\bibitem[{{World Health Organization}(2018)}]{WHO2018}
\bibinfo{author}{{World Health Organization}}, \bibinfo{year}{2018}.
\newblock \bibinfo{title}{Global action plan on physical activity 2018-2030: more active people for a healthier world}.
\newblock \URLprefix \url{https://www.who.int/publications/i/item/9789241514187}. \bibinfo{note}{accessed: 2024-08-14}.

\end{thebibliography}

\end{document}